\documentclass[review, 3p, 12pt]{elsarticle}
\usepackage{soul}
\usepackage{graphicx, graphics}
\usepackage{syntonly}
\usepackage{amssymb, amsmath, amsfonts,nccmath}
\usepackage{subfigure}
\usepackage{multirow}
\usepackage{color}
\usepackage{hyperref}

\usepackage{bm}

\providecommand{\pfrac}[2]{\frac{\partial #1}{\partial #2}}

\providecommand{\bme}{\bm \varepsilon}
\providecommand{\bms}{\bm \sigma}

\providecommand{\dbme}{\dot{\bm \varepsilon}}
\providecommand{\dbms}{\dot{\bm \sigma}}

\begin{document}
% Title page details:
\title{Bond-based nonlocal models by nonlocal operator method in symmetric support domain}
%\author{Huilong Ren, Xiaoying Zhuang, Timon Rabczuk}
\author[HU]{Huilong Ren}
\ead{huilong.ren@iop.uni-hannover.de}
\author[HU,TU]{Xiaoying Zhuang\corref{cor1}}
\ead{zhuang@iop.uni-hannover.de }
\author[Hu2]{Xiaolong Fu}
%\ead{zhuang@iop.uni-hannover.de }
\author[Hu1]{Zhiyuan Li}
\ead{lizhiyuan1007@163.com}
\author[BU]{Timon Rabczuk}
\ead{timon.rabczuk@uni-weimar.de}

%\address[BU]{IOP, Leibniz University Hannover}
\address[TU]{State Key Laboratory of Disaster Reduction in Civil Engineering, College of Civil Engineering,Tongji University, Shanghai 200092, China}
\address[HU]{Institute of Photonics, Department of Mathematics and Physics, Leibniz University Hannover,Germany}
\address[Hu2]{Xi’an Modern Chemistry Research Institute, Xi’an, 710065, China}
\address[Hu1]{Department of Engineering Mechanics, Hohai University, Nanjing, 211100, China}
\address[BU]{Institute of Structural Mechanics, Bauhaus-Universit at Weimar, Germany}
\cortext[cor1]{Institute of Photonics, Department of Mathematics and Physics, Leibniz University Hannover,Germany. zhuang@iop.uni-hannover.de;}
\begin{abstract}
The present study focuses on the applications of energy decomposition in diverse nonlocal models, such as elasticity, thin plates, and gradient elasticity, with the aim of establishing bond-based nonlocal models in which the bond force is solely dependent on the deformation of a single bond. Through the adoption of an appropriate bond force form and the application of energy equivalence between local and nonlocal models, several kinds of highly succinct bond-based models are obtained. The present study involves a reexamination of nonlocal operator methods, with a particular focus on the simplified version within a symmetric support domain. A three-point bent-bond model has been proposed to characterize the curvature and bending moment. A crack criterion for normal strain of the bond based on Griffith theories is proposed. This approach is analogous to the phase field model and allows for individual application to each bond, resulting in strain localization. By implementing this rule, the path of the crack can be predicted in an automated manner through the act of cutting the bond, yielding outcomes that are akin to those obtained via the phase field method. Simultaneously, a crack rule for critical shear strains in shear fractures is presented. Moreover, an incremental version of the plasticity model associated with bond force has been formulated. The nonlocal bond-based models are further validated through several numerical examples.

\end{abstract}
\maketitle

\section{Introduction}
The issue of material damage and structural failure persists in various engineering applications. Insufficient depiction may result in significant risks and economic damages. The issue stems from the inherent difficulty in predicting the intricate mechanisms involved in the process of damage through theoretical models and numerical methods. In recent decades, significant efforts have been devoted to the development of robust numerical methods, such as damage mechanics \cite{de2016gradient,areias2016damage}, phase field method \cite{miehe2010thermodynamically,borden2012phase}, extended finite element method \cite{sukumar2000extended,moes2002extended}, meshless methods \cite{belytschko1994element,liu1995reproducing}, cracking particle method \cite{rabczuk2004cracking,rabczuk2010simple}, virtual crack closure technique\cite{krueger2004virtual}, Peridynamics (PD) \cite{Silling2007}, and several others. Typically, these techniques can be classified into two distinct groups: One approach is to incorporate an auxiliary field to depict the presence of a crack or to alter the topology of the material to generate a surface that represents the crack. The incorporation of an auxiliary field in the phase field method facilitates the determination of the crack surface topology while preserving the integrity of the mesh. This approach is characterized by its numerical stability and smoothness, albeit at the cost of solving an additional field. The approach utilizing topology modification has the capability to generate a well-defined crack surface, although the utilization of geometric manipulation may result in instability issues. The aforementioned categories appear to possess distinct characteristics, however, they can both be attributed to Griffith theory. This theory posits that the emergence of new free surfaces is a result of the transformation of diminished strain potential energy into surface energy \cite{griffith1921vi}.

Phase field methods and nonlocal methods are two prominent examples of fracture modeling. The phase field approach has the capability to address various complex engineering fracture problems within a relatively simple theoretical framework, as evidenced by the works of Wu et al. \cite{wu2020phase}, Mikelic et al \cite{mikelic2015phase} , Msekh et al. \cite{msekh2015abaqus}, Amiri et al.\cite{amiri2014phase}, Zhou et al. \cite{zhou2020phase}, and Dittmann et al. \cite{dittmann2018variational}. The nonlocal theory of Peridynamics offers certain benefits in the realm of topology modification, owing to its ability to independently consider interactions within a domain of finite size. Stated differently, the density of strain energy in PD is dispersed throughout the given domain, as opposed to being concentrated at a point lacking dimensions. The act of breaking individual bonds can be better understood through physical intuition, particularly in the context of cracking. The literature presents several examples of peridynamic models, such as bond-based PD \cite{silling2000reformulation}, bond-based PD incorporating shear deformation \cite{ren2016new}, extended bond-based PD \cite{zhu2017peridynamic,ni2019static,madenci2021bond}, conjugate bond pair-based PD \cite{wang20183}, bond-based micropolar PD \cite{diana2019bond}, among others. The bond-based peridynamic approach exhibits favorable numerical stability in fracture modeling due to the splitting of energy among individual bonds. {An intriguing extension of bond-based PD theory is the continuum-kinematics-inspired peridynamics, as proposed by Javili et al. }\cite{javili2019continuum}. {This approach incorporates two- or three-neighbor interactions, building upon the bond-based PD. The two- or three-neighbors interaction can be cut down similarly to the bond-based PD, as these interactions are mutually exclusive. This feature may offer advantages for fracture modeling.} The state-based PD  proposed by Silling \cite{Silling2007}, has the capability to address continuous problems. However, the stability of the bond cutting process is compromised due to the complete coupling of all bonds within the horizon. The nonlocal operator method (NOM) has been proposed in Ref. \cite{ren2020nonlocal,rabczuk2019nonlocal,ren2020higher} as an extension of the dual-horizon PD \cite{Ren2015}. The authors have proposed a methodology that offers a systematic approach to convert numerous local models into their nonlocal counterparts. Additionally, they have introduced a variational framework that can be employed to tackle challenging problems \cite{ren2020nonlocalgrad,Ren2021May,ren2021nonlocal}.

Bond-based models offer significant versatility in the realm of fracture modeling. The concept of NOM pertains to the computation of the function's derivatives with respect to the overall information contained in its support. The constancy or regularity of shapes in the support domain is not a prerequisite. The general structure of the NOM appears intricate, and specifically, it is not directly applicable for simulating fracture through bond cutting. Given the benefits of bond-based peridynamics, including its capacity for automatic crack development and determination of crack direction, it is our aim to devise more general bond-based models for diverse mechanical problems, thereby enabling the utilization of this attribute for fracture modeling. Fracture creation through bond cutting relies on the critical strain present in bond-based PD. The congruity of the precise critical strain derived from Griffith theory within nonlocal models remains incongruous in relation to load-displacement curves when compared to alternative methodologies, such as the phase field method. Consequently, the research aims to accomplish two primary objectives. Firstly, to formulate bond-based nonlocal models utilizing the nonlocal operator method in a symmetric support domain. Secondly, to ascertain a suitable critical stretch value based on Griffith's theory. The study is limited to the setting of small deformation for the sake of simplicity.

The subsequent sections of this writing are structured in the following manner. Section 2 provides a succinct overview of the nonlocal operator method and delves into the basic principles of local and nonlocal models. The third section expounds upon the orthogonal decomposition of energy as it pertains to fracture modeling using both the phase field method and bond-based peridynamics. In Section 4, a second-order NOM with symmetric support is presented in a simplified form. This section presents the derivation of the weighted bond-based nonlocal bar and nonlocal beam model. Section 5 of the paper presents a detailed derivation of the bond-based nonlocal elasticity in both two-dimensional and three-dimensional settings, utilizing energy equivalence. A bond-cutting criterion that is based on normal or shear strains is proposed as a simple yet effective approach, drawing an analogy to the phase field method. The derivation of the plastic model for the bond in nonlocal elasticity is presented. Section 6 presents the derivation of the nonlocal bond-based isotropic thin plate model and the nonlocal bond-based gradient elasticity using the second-order NOM under symmetric support. Section 7 outlines three numerical experiments, namely the nonlocal simply supported beam, crack propagation in a single-edge-notched plate under tension/shear boundary conditions, and the Kalthoff-Winkler test with tension and shear fractures. Section 8 presents conclusions and an outlook.

\section{Review of nonlocal operator method}
Following the notations in Ref \cite{ren2020nonlocal,ren2020higher,ren2021nonlocal}, we briefly outline the results of NOM. NOM uses the integral form to replace the partial differential derivatives of different orders based on the concept of support and dual-support.
\subsection{Support and dual-support}\label{sec:nom}
\begin{figure}[htp]
\centering
\includegraphics[width=.4\textwidth]{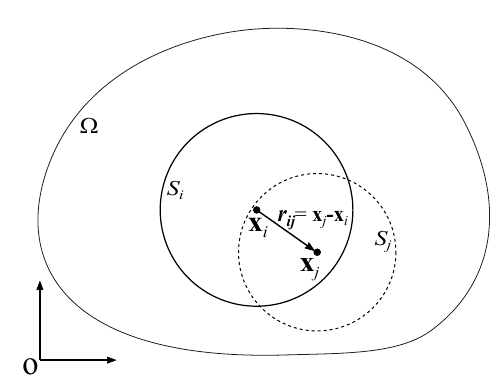}
\caption{Support and bond in NOM.}\label{fig:Coord0}
\end{figure}
Consider a domain as shown in Figure \ref{fig:Coord0}, let $\bm x_i$ be spatial coordinates in the domain $\bm \Omega$; $\bm r_{ij}:=\bm x_j-\bm x_i$ is a spatial vector starting from $\bm x_i$ to $\bm x_j$; $ u_i:= u(\bm x_i,t)$ and $ u_j:= u(\bm x_j,t)$ are the field values for $\bm x_i$ and $\bm x_j$, respectively; $ u_{ij}:= u_j- u_i$ is the relative field for spatial vector $\bm r_{ij}$.

\textbf{Support} $\mathcal S_{i}$ is the neighbourhood of point $\bm x_i$. A point $\bm x_j$ in support $\mathcal S_i$ forms the spatial vector $\bm r_{ij}(=\bm x_j-\bm x_i)$. The support in NOM can be a spherical domain, a cube, semi-spherical domain and many others. In a spherical domain, the radius of support $\mathcal S_i$ is represented by $\delta_i$. 

\textbf{Dual-support} is defined as a union of points whose supports include $\bm x_i$, denoted by
\begin{align}
\mathcal S_{i}'=\{\bm x_j|\bm x_i \in \mathcal S_{j}\}. \notag%\label{eq:dualsupport}.
\end{align}
Point $\bm x_j$ forms the dual-vector $\bm r_{ji}(=\bm x_i-\bm x_j=-\bm r_{ij})$ in $\mathcal S_{i}'$. On the other hand, $\bm r_{ji}$ is the spatial vector formed in $\mathcal S_{j}$. The terms $\bm r_{ij}$ and $\bm r_{ji}$ are also used to refer to bonds in nonlocal theories. It is worth mentioning that the size of the support of each point can be different. When the support sizes for all material points are the same, the dual-support is equal to the support.

In two-dimensinal space, the second-order nonlocal derivatives of field $u$ in support $\mathcal S_i$ \cite{ren2021nonlocal} are calculated as
\begin{subequations}
\begin{align}
\tilde{\partial} u_{i}&=\left(u_{i, x}, u_{i, y}, u_{i, x x}, u_{i, x y}, u_{i, y y}\right)^{T} :=\int_{\mathcal{S}_{i}} \omega({r}_{i j}) \bm{K}_{i} \cdot \bm{p}_{i j} u_{i j} d V_{j} \label{eq:partialui}\\
\mbox{with}\quad u_{i j} &=u_{j}-u_{i}=u\langle \bm r_{ij} \rangle , \bm r_{ij}=\bm r=(r_x,r_y)=(x_{ij},y_{ij})=r \bm n\\
\bm{p}_{i j} &=\left(x_{i j}, y_{i j}, x_{i j}^{2} / 2, x_{i j} y_{i j}, y_{i j}^{2} / 2\right)^{T} \\
\bm{K}_{i} &=\left(\int_{\mathcal S_{i}} \omega({r}_{i j}) \bm{p}_{i j} \otimes \bm{p}_{i j}^{T} d V_{j}\right)^{-1},\label{eq:Ki}
\end{align}
\end{subequations}
where $\tilde{\partial} u_{i}$ represents a collection of nonlocal derivatives, $\omega(r)$ denotes the weight function, $\bm n$ signifies the unit direction of bond $\bm r$, $r$ represents the magnitude of $\bm r$, $x_{ij}=x_j-x_i$ and the superscript $T$ denotes the transposition of a vector or a matrix. It is noteworthy that $\omega(r)$ has the potential to adopt diverse expressions, for example, $\omega(r)=1$ or $\omega(r)=1/r^2$. The symbols $\square_{i j}$ and $\square\langle \bm r \rangle$ are utilized in the current context to denote the physical parameters that are linked to a bond. The symbol $\square\langle \bm r \rangle$ is employed in situations where the bond pair is not explicitly designated. 

Let $\bm K_i \cdot \bm p_{ij}$ be denoted by
$
(g_{1j},g_{2j},h_{1j},h_{2j},h_{3j})^T= \bm K_i \cdot \bm p_{ij}
$.
 The gradient vector $\bm g_{ij}$ and Hessian matrix $\bm h_{ij}$ between points $i$ and $j$ in 2D can be constructed as, respectively
\begin{align}
\bm g_{ij}=(g_{1j},g_{2j})^T,\quad
\bm h_{ij}=\begin{pmatrix}
h_{1j} & h_{2j}\\
h_{2j} & h_{3j}\\
\end{pmatrix}.\label{eq:ghgeneral}
\end{align}

Accordingly, the nonlocal gradient operator and Hessian operator for vector field (e.g. $\bm u$) can be defined as
\begin{subequations}
\begin{align}
%\begin{subequations}
\tilde{\nabla} \otimes \bm u_i&:=\int_{\mathcal S_i} \omega(r_{ij})\bm u_{ij} \otimes \bm g_{ij} \, d V_j\label{eq:nu1}\\
\tilde{\nabla}\otimes \tilde{\nabla} \otimes \bm u_i&:=\int_{\mathcal S_i} \omega(r_{ij})\bm u_{ij} \otimes \bm h_{ij} \, d V_j.\label{eq:nu2}
%\end{subequations}
\end{align}
\end{subequations}
In above equations, we use $\tilde{\nabla}$ to denote the nonlocal version of gradient. More details of the derivation can be found in Ref \cite{ren2021nonlocal}.

\subsection{Variational derivation of nonlocal models}
The nonlocal model can be obtained from the local energy functional through the utilization of either the nonlocal gradient or nonlocal Hessian, as is typical in NOM \cite{ren2021nonlocal}. In the context of a general gradient elastic solid, where the energy density $\phi$ is dependent on the gradients of displacement field denoted by $\bm u$ and its Laplacian, the overall internal potential energy $\Psi$ within the domain $V$ can be expressed as follows
\begin{align}
\Psi=\int_V \phi(\nabla \bm u,\nabla^2\bm u) d V.\notag
\end{align}
The variation of the energy functional
\begin{align}
\delta\Psi&=\int_V\frac{\partial\phi}{\partial \nabla \bm u}:\nabla \delta \bm u+\frac{\partial\phi}{\partial \nabla^2 \bm u}\dot{:}\nabla^2 \delta \bm u d V\notag\\
&=\int_V\bm \sigma_i:\int_{\mathcal S_i} \omega(r_{ij}) \delta\bm u_{ij} \otimes \bm g_{ij}d V_j+\bm \Sigma_i \dot{:}\int_{\mathcal S_i} \omega(r_{ij}) \delta\bm u_{ij} \otimes \bm h_{ij} d V_j dV_i\notag\\
&=\int_V \int_{\mathcal S_i} \omega(r_{ij})\bm \sigma_i\cdot \bm g_{ij} \cdot \delta \bm u_{ij} d V_j +\int_{\mathcal S_i} \omega(r_{ij})\bm \Sigma_i :\bm h_{ij} \cdot \delta \bm u_{ij}  d V_j dV_i,\label{eq:varDeriNOM}
\end{align}
where $\bm \sigma:=\frac{\partial\phi}{\partial \bm \varepsilon}$ is the stress tensor, $\bm \varepsilon=\frac 12 (\nabla \bm u+\nabla \bm u^T)$ is the strain tensor described by displacement gradient $\nabla\bm u$, and $\bm \Sigma:=\frac{\partial\phi}{\partial \nabla \bm\varepsilon}$ is the couple stress. In the derivation of Equation \ref{eq:varDeriNOM}, the nonlocal gradient (Hessian) by Equation \ref{eq:nu1} are used to replace the local gradient (Hessian). 
For the cases of the linear elasticity and the linear gradient elasticity, the material constitutions are
\begin{align}
\bm \sigma=\mathbb C:\bm \varepsilon, \mbox{ or } \sigma_{ij}=C_{ijkl} \varepsilon_{kl}\notag\\
\bm \Sigma=\mathbb D:\nabla\bm \varepsilon, \mbox{ or } \Sigma_{ijk}=D_{ijklmn} \partial_l \varepsilon_{mn},\notag
\end{align}
where $\mathbb C$, $\mathbb D$ are material tensors and $C_{ijkl}, D_{ijklmn}$ are entries of material tensors.

With some mathematical manipulation, the nonlocal governing equations for elasticity and gradient elasticity are, respectively, 
\begin{subequations}
\begin{align}
\int_{\mathcal S_i} \omega(r_{ij})\bm \sigma_i \cdot \bm g_{ij}d V_j&-\int_{\mathcal S_i'}\omega(r_{ji})\bm \sigma_j \cdot \bm g_{ji}d V_j+\bm b=\rho \ddot{\bm u}_i,\label{eq:nlelas}\\
\int_{\mathcal S_i} \omega(r_{ij})(\bm \sigma_i \cdot \bm g_{ij}+\bm \Sigma_i:\bm h_{ij})d V_j&-\int_{\mathcal S_i'}\omega(r_{ji})(\bm \sigma_j \cdot \bm g_{ji}+\bm \Sigma_j:\bm h_{ji})d V_j+\bm b=\rho \ddot{\bm u}_i,\label{eq:nlgrade}
\end{align}
\end{subequations}
where $\bm b$ denotes the body force density, $\rho$ the density and $\ddot{\bm u}_i$ the acceleration.

%Based on the material constitutive of thin plate
%\begin{align}
%\bm{M}=\left(\begin{array}{cc}
%M_{x x} & M_{x y} \\
%M_{x y} & M_{y y}
%\end{array}\right)
%=D_{0}\left(\nu \mbox{tr}(\bm{\kappa}) \bm{I}_{2 \times 2}+(1-\nu) \bm{\kappa}\right),
%\end{align}
%the nonlocal governing equation of nonlocal thin plate is
%\begin{align}
%\int_{\mathcal S_i} ( f_{ij}+ f_{ji})+q=\rho \ddot{w}_i, \mbox{ with }f_{ij}=\omega(r_{ij})\bm M_{ij}:\bm h_{ij}\label{eq:nlthinp},
%\end{align}
%where $D_{0}=\frac{E t^{3}}{12\left(1-\nu^{2}\right)}$ and $t$ is the thickness of the plate.
 The detailed derivation of nonlocal governing equations in Equations \ref{eq:nlelas},\ref{eq:nlgrade} can be found in Ref  \cite{ren2021nonlocal}. The equation presented above is dependent upon the state quantity that has been defined within the support domain. It is important to note that in this context, the internal force of each bond is fully coupled. 

\subsection{Local form vs nonlocal form}
The formulation of the local model is based on differential equations, while nonlocal theories, such as peridynamics, are represented through integral equations. Mathematically, the local form and nonlocal form of balance equation of elasticity can be written as, respectively
\begin{align}
\nabla \cdot \bm \sigma+\bm b=\rho\ddot{\bm u},\quad \textstyle{\int}_{\mathcal S_i} (\bm f_{ij}-\bm f_{ji}) d V_j+\bm b=\rho\ddot{\bm u}_i.\notag
\end{align}
Herein, the constant support for all material points is assumed and $\bm f_{ij}$ denotes the bond force density. 

\begin{figure}[htp]
\centering
\includegraphics[width=12cm]{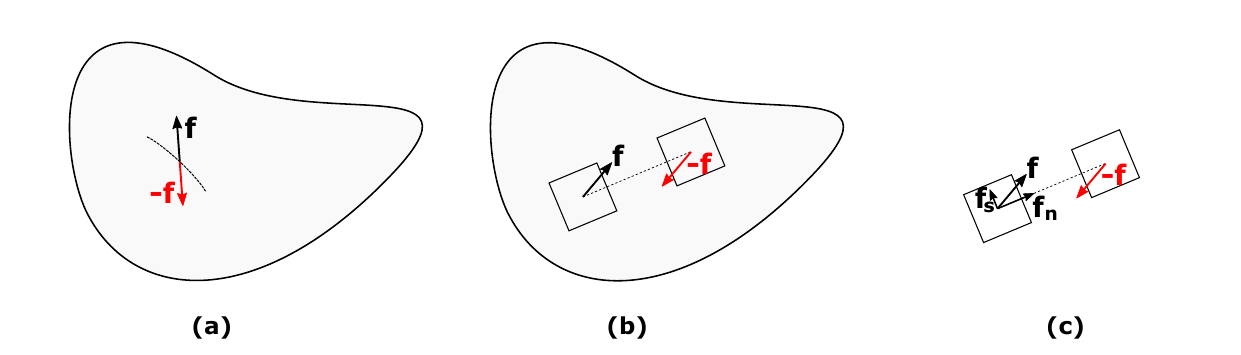}
\caption{Internal forces (a) between virtual segments in local theory, (b) between two micro-volumes in nonlocal theory. (c) bond-force decomposition $\bm f=\bm f_s+\bm f_n$.}
\label{fig:bondforces0}
\end{figure}

The primary disparity between the local and nonlocal models lies in their respective approaches to internal force, as depicted in Figure \ref{fig:bondforces0}. {The former is characterized by its reliance on point structure, devoid of any consideration for shape or size. For any point on the line segment, there is a pair-wised force $(\bm f, -\bm f)$ with opposite directions, as shown in Figure} \ref{fig:bondforces0}(a). In contrast, the latter incorporates a finite-distance neighborhood to explicitly account for nearby interactions. In Figure \ref{fig:bondforces0}(b), any two micro-volumes of finite distance form a pair, which results in internal force $\bm f$. In addition, the bond force $\bm f$ can be decomposed into the normal force $\bm f_n$ along the bond direction and the shear force $\bm f_s$ perpendicular to the bond direction, as depicted in Figure \ref{fig:bondforces0}(c).
\begin{figure}[htp]
\centering
\includegraphics[width=6cm]{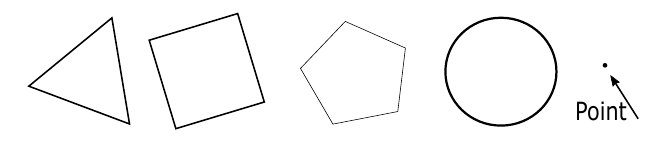}
\caption{Shapes of support.}
\label{fig:bondforces}
\end{figure}
Dual-horizon PD allows for significant flexibility in the shape of the horizon, although circular horizons are typically preferred. The finite size horizon domain allows for the distribution of internal force while maintaining the same physical quantities, such as density, as in conventional local theory. This provides an advantageous means of manipulating internal forces in the event of discontinuity or damage. Local theories that are conventional in nature are formulated based on sets of points. From a mathematical standpoint, a point is a geometric object that possesses no discernible shape and is infinitely small. Splitting a single point into two separate points can be considered an inconvenience. When considering the ideal symmetry of a geometric object, the circular domain in 2D or spherical domain in 3D is the shape that is closest to a point. %

The preceding subsection has demonstrated that nonlocal forms can be established on arbitrary support settings through the variational derivation of the energy functional. It has been noted that these models are classified as state-based nonlocal models. Like the state-based peridynamics, the removal of neighbors from the support in this method can lead to singularity, thereby limiting its applicability in modeling discontinuities such as fractures.  It is worth noting that in cases where the support domain is completely symmetric, nonlocal models can be significantly simplified, particularly in relation to nonlocal operator methods. It is feasible to formulate bond-based variations of nonlocal models through this approach.
%Another consideration 
%

\section{Energy orthogonal decomposition for fracture evolution}
Fracture modeling encounters challenges in state-based nonlocal models due to the interdependence of bonds in support, which are linked through the state defined by all bonds. It can be argued that the deformation energy ought to be considered as a whole. By way of comparison, the phase field method and the bond-based PD are two effective techniques for modeling fractures. These methods exhibit remarkable numerical stability when dealing with challenging fracture problems. The rationale behind this is that they rely on the energy orthogonal decomposition of the strain energy density. 
\subsection{Phase field model}
Miehe \cite{miehe2010thermodynamically} proposed a thermodynamic consistent phase field model for brittle fracture. The success of this model lies in the orthogonal decomposition of strain energy density (e.g. $\psi_e$) of isotropic linear elasticity
\begin{align}
\psi_e=\frac 12\bm \sigma:\bm\varepsilon=\frac 12(\sum_{i=1}^3\sigma_i \bm n_i \otimes \bm n_i):(\sum_{i=1}^3\varepsilon_i \bm n_i \otimes \bm n_i)=\frac 12{\sum}_{i=1}^3 \sigma_i \varepsilon_i,\notag
\end{align}
where $\sigma_i$ is the $i$-th principal stress on direction $\bm n_i$, ($\bm n_1,\bm n_2,\bm n_3$) are the eigenvectors of the associated principal strains ($\varepsilon_1,\varepsilon_2,\varepsilon_3$) of $\bm\varepsilon$, the stress tensor $\bm \sigma=\sum_{i=1}^3\sigma_i \bm n_i \otimes \bm n_i$ and strain tensor $\bm \varepsilon=\sum_{i=1}^3\varepsilon_i \bm n_i \otimes \bm n_i$ are formulated based on the eigenvalue decomposition of a matrix.
For linear isotropic elasticity, the orthogonal decomposition of stress tensor and strain tensor are co-axial and the positive/negative parts of stress tensor can be written as
\begin{align}
\bm\sigma^{\pm}=&\frac{\partial\psi_e^{\pm}}{\partial\bm\varepsilon}=\lambda\langle\varepsilon_1+\varepsilon_2+\varepsilon_3\rangle_{{\pm}}\bm I \notag\\
&+2\mu (\langle\varepsilon_1\rangle_{{\pm}}\bm n_1\otimes\bm n_1+\langle\varepsilon_2\rangle_{{\pm}}\bm n_2\otimes\bm n_2+\langle\varepsilon_3\rangle_{{\pm}}\bm n_3\otimes\bm n_3),\notag%\label{eq:MieheSP}
\end{align}
where $\bm I$ is the identity matrix, ($\lambda,\mu$) denote the Lame constants, $\langle x\rangle_{{\pm}}=(x\pm |x|)/2$ , and $\psi_e^{\pm}=\frac 12\lambda\langle\varepsilon_1+\varepsilon_2+\varepsilon_3\rangle_{{\pm}}^2+\mu (\langle\varepsilon_1\rangle_{{\pm}}^2+\langle\varepsilon_2\rangle_{{\pm}}^2+\langle\varepsilon_3\rangle_{{\pm}}^2)$. {The utilization of the orthogonal property results in a reduction of the number of terms present in the mathematical expression of strain energy from 9 to 3, thereby significantly simplifying the aforementioned expression.  A drawback associated with the phase field model pertains to the intricate computation of partial derivatives of eigenvalues and eigenvectors in relation to strain tensors.}

\subsection{Bond-based PD}

One of the most successful Peridynamics is the bond-based version. In bond-based PD, the internal potential energy density carried by a point is
\begin{align}
\psi=\frac 12\int_{\mathcal S} c s^2 d V,\notag
\end{align}
where variable $s$ denotes the extensional strain pertaining to the bond, while $c$ represents the material parameter. {Herein, we restrict the discussion of bond-based PD in the setting of small deformation for simplicity.} The energy associated with each bond is considered to be independent of each other. Despite the Poisson's ratio restriction, the original bond-based PD model is highly stable when it comes to simulating tensile fractures. %In constrast to the state-based PD, whose  bond-force is contingent upon the state variables, such as the stress tensor. The breakage of a bond results in the singularity of the shape tensor. The singularity phenomenon is absent in bond-based PD due to the autonomy of individual bonds. 

The models of local elasticity and nonlocal elasticity are interconnected through the principle of energy equivalence. It is possible to obtain numerous nonlocal models through variational derivations based on local models \cite{Ren2021Sep},. The present discussion centers on linear elasticity, wherein the strain energy density is expressed as follows:
 $\bm \varepsilon :\mathbb C:\bm \varepsilon$, where $\bm \varepsilon$ is strain tensor. Then we use nonlocal gradient to replace strain tensor, e.g. $\bm\varepsilon\to \textstyle{\int}_{\mathcal S_i} \omega(r_{ij}) \bm u_{ij} \otimes \bm g_{ij} dV_j$, where $\bm u_{ij}=\bm u_j-\bm u_i$,  the equivalent nonlocal energy density can be conceptually written as
 \begin{subequations}
\begin{align}
&(\textstyle{\int}_{\mathcal S_i} \omega(r_{ij}) \bm u_{ij} \otimes \bm g_{ij} dV):\mathbb C: (\int_{\mathcal S_i} \omega(r_{ij}) \bm u_{ij} \otimes \bm g_{ij} dV) \label{eq:twoinifinite}\\
&\overset{?}{=} \textstyle{\int}_{\mathcal S_i} (\omega(r_{ij}) \bm u_{ij} \otimes \bm g_{ij}):\bar{\mathbb C} (\bm r_{ij}): ( \bm u_{ij} \otimes \bm g_{ij}) dV ,\label{eq:oneinifinite}
\end{align}
\end{subequations}
where $\bar{\mathbb C} (\bm r_{ij})$ is the material tensor for a single bond $\bm r_{ij}$. Each integral form comprises an infinite number of terms, and the product of two integral forms results in a greater number of infinite terms. It would be highly desirable for Equation \ref{eq:oneinifinite} to be equivalent to Equation \ref{eq:twoinifinite}. The aforementioned proposition pertains to the determination of whether the expression $(\sum_{i=1}^n \bm a_i)\cdot (\sum_{i=1}^n \bm b_i)$ is equivalent to $\sum_{i=1}^n \bm a_i\cdot \bm b_i$. The derivation necessitates the fulfillment of the orthogonal condition, which is expressed as $\bm a_i \cdot \bm b_j=\delta_{ij}$ in mathematical terms.

\section{Nonlocal operator method in symmetric support}
NOM formulated on general support domain does not satisfy the orthognoal conditions. Consider the advantage of bond-based PD in symmetric horizon, it is reasonable to streamline the concept of NOM through the adoption of a symmetric support. This section will analyze the NOM within the context of a symmetric support domain. %Additionally, the process of obtaining bond-based elastic solids, thin plates, and gradient solid is explored, wherein the reaction force and direct force need not be taken into account for each bond. The independence of the internal force calculation for each material point presents a noteworthy advantage for its practical application.
\subsection{NOM in 2D/3D symmetric support}
In the first-order NOM or PD, for circular or spherical support, the shape tensor can be written by the identity matrix
\begin{align}
\int_{\mathcal S} \omega(r) \bm r\otimes \bm r d V=\alpha_k \bm I\notag
\end{align}
with coefficients defined as $\alpha_{2}=\int_{0}^\delta \omega(r) \pi r^3 d r$ in 2D, $\alpha_{3}=\int_0^\delta \omega(r) \frac 43 \pi r^4 d r$ in 3D.

The bond-based gradient vector $\bm g$ in 2D/3D can be uniformly written as
\begin{align}
\bm g=\frac{r \bm n}{\alpha_k},\label{eq:g2d}
\end{align}
where $k\in \{2,3\}$ and $\bm n=(n_x,n_y)=\bm r/r$ in 2D or $\bm n=(n_x,n_y, n_z)=\bm r/r$ in 3D. 

The nonlocal gradient, nonlocal divergence and nonlocal curl operator using explicit bond notations can be rewritten as
\begin{align}
\tilde{\nabla} * \bm u_i:=\int_{\mathcal S_i} \omega(r_{ij}) \bm g_{ij} * \bm u_{ij} d V_j,\notag
\end{align}
where $*\in \{\otimes,\cdot, \times\}$ and $\bm g_{ij}$ is defined in Equation \ref{eq:g2d} for bond $\bm r_{ij}$.

In case of second-order NOM with support domain taking the form of a circular area, the formulation of $\bm K_i$ in Equation \ref{eq:Ki} can be significantly streamlined. Through mathematical manipulation, the nonlocal gradient and nonlocal Hessian in Equation \ref{eq:partialui} can be simplified as
\begin{align}
\tilde{\partial}u=\int_{\mathcal S} \omega(r) u\langle \bm r \rangle\big(\frac{(r_x,r_y)}{\pi \int_0^{\delta } \omega (r) r^3 \, dr},\frac{1}{\pi \int_0^{\delta } \omega (r) r^5 \, dr}(3 r_x^2-r_y^2, 4 r_x r_y, 3 r_y^2-r_x^2)\big) d V.\notag%\label{eq:nomSimple}
\end{align}
The nonlocal derivative contribution of a single bond can be expressed in a straightforward manner as
\begin{align}
\tilde{\partial}u\langle \bm r \rangle=\omega(r) u\langle \bm r \rangle\big(\underbrace{\frac{(r_x,r_y)}{\pi \int_0^{\delta } \omega (r) r^3 \, dr}}_{\bm g},\underbrace{\frac{1}{\pi \int_0^{\delta } \omega (r) r^5 \, dr}(3 r_x^2-r_y^2, 4 r_x r_y, 3 r_y^2-r_x^2)}_{\vec{\bm h}}\big)\label{eq:partialur},
\end{align}
where the gradient vector $\bm g=\frac{(r_x,r_y)}{\pi \int_0^{\delta } \omega (r) r^3 \, dr}$ and vector $\vec{\bm h}=\frac{1}{\pi \int_0^{\delta } \omega (r) r^5 \, dr}(3 r_x^2-r_y^2, 4 r_x r_y, 3 r_y^2-r_x^2)$. In contrast to the general form of gradient vector and Hessian matrix in Equation \ref{eq:ghgeneral}, the adoption of symmetric support allows an explicit expressions of $\bm g$ and $\vec{\bm h}$.

Based on Equation \ref{eq:partialur}, we extract bond curvature tensor $\bm h^{2d}$ from vector $\vec{\bm h}$ in 2D, which is written as
\begin{align}
\bm h^{2d}=\frac{r^2}{\pi \int_0^{\delta } \omega (r) r^5 \, dr}\begin{pmatrix}
4 n_x^2-1 & 4 n_x n_y\\
4 n_x n_y & 4 n_y^2-1\\
\end{pmatrix}=\frac{r^2}{\pi \int_0^{\delta } \omega (r) r^5 \, dr}\big(4 \bm n\otimes \bm n-\bm I\big).\label{eq:h2d}
\end{align}
Similarly, the matrix form of the bond curvature in 3D is
\begin{align}
\bm h^{3d}
&=\frac{3 r^2 }{4 \pi \int_0^{\delta } \omega (r) r^6 \, dr}
\big(5 \bm n\otimes \bm n-\bm I\big).
\end{align}

%\begin{frame}{Round/spherical support}
%For the case of gradient operator, the nonloal gradient of a vector field can also be written as
%\begin{align}
%\nabla \bm u=\int_{\mathcal S} \frac{3 \omega(r) }{4 \pi \int_0^{\delta } \omega (r) r^4 \, dr}(r_x,r_y,r_z)\otimes \bm u\langle \bm r \rangle dV=\int_{\mathcal S}\omega(r) \bm g\otimes \bm u\langle \bm r\rangle d V.\label{eq:nongradu}
%\end{align}

\begin{figure}[htp]
\centering
\includegraphics[width=4cm]{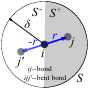}
\caption{Bond and bent bond. $\mathcal S=\mathcal S^+\cup \mathcal S^-$. $j'\in \mathcal S^-$ is the symmetric point of $j\in \mathcal S^+$ with respect to center point $i$, $ijj'$ for a bent bond.}
\label{fig:pairbond}
\end{figure}

The Hessian of a scalar field in 2D has the form
\begin{align}
\nabla \nabla u&=\int_{\mathcal S}\frac{\omega(r)u\langle \bm r \rangle}{\pi \int_0^{\delta } \omega (r) r^5 \, dr}\begin{pmatrix}
3 r_x^2-r_y^2 & 4 r_x r_y\\
4 r_x r_y & 3 r_y^2-r_x^2\\
\end{pmatrix}dV\notag\\
&=\int_{\mathcal S}\omega(r)\bm h\otimes u\langle \bm r\rangle d V\notag\\%\label{eq:ddu-1}\\
&=\int_{\mathcal S^+}\omega(r)\bm h\otimes( u\langle \bm r\rangle+ u\langle -\bm r\rangle) d V.\label{eq:ddu}
\end{align}
In Equation \ref{eq:ddu}, $\bm h$ is invariant for both $\bm r$ and $-\bm r$ and the symmetry of $\mathcal S$ is considered. The half support $\mathcal S^+$ is defined based on the symmetric support domain as shown in Figure \ref{fig:pairbond}. {The conventional bond, denoted as $\bm r$, is characterized by a requirement of solely two points $i$ and $j$. Upon consideration of symmetry, it can be observed that the mirror image point $j'$ of point $j$, when combined with points $i$ and $j$, results in the formation of a bent bond that is denoted by $ijj'$. The definition of bond curvature requires only a half support.}
{It is worth mentioning that the bent bond exhibits similarities to the two-neighbor interaction} \cite{javili2019continuum}. {The concept of a bent bond pertains to the curvature of symmetric material points within a support, whereas the two-neighbor interaction for elastic solids can be established between any two material points in the horizon.}

Traditional NOM deals with the gradient or Hessian at a point as a whole. In this sense, all bonds in support domain are coupled. In the spirit of bond-based PD, it is natural to define the derivatives for each individual bond. Based on Equation \ref{eq:g2d}, the bond gradient on single bond is defined as
\begin{align}
\nabla  u\langle \bm r\rangle= \frac{\bm r}{r^2} u \langle \bm r\rangle=u\langle\bm r\rangle \frac{\bm n}{r}.\notag%\label{eq:bondgrad}
\end{align}
Based on Equation \ref{eq:h2d}, the curvature of a pair-wised bond in 2D is defined as
\begin{align}
\nabla \nabla u\langle \bm r\rangle=\frac{ (u\langle \bm r\rangle+u\langle -\bm r\rangle)}{r^2} (4 \bm n\otimes\bm n-\bm I).\label{eq:dd2bond}
\end{align}
Above definition is reasonable because $\frac{ u\langle \bm r\rangle}{r^2} (4 \bm n\otimes\bm n-\bm I)\approx (\nabla \nabla u:(\bm r\otimes \bm r))/r^2 (4 \bm n\otimes\bm n-\bm I)=(\nabla \nabla u:(\bm n\otimes \bm n))(4 \bm n\otimes\bm n-\bm I)$, which depends on the bond direction and second-order derivatives.

Similarly, the curvature of a pair-wised bond in 3D can be written as
\begin{align}
\nabla \nabla u\langle \bm r\rangle=\frac{ (u\langle \bm r\rangle+u\langle -\bm r\rangle)}{r^2} (5 \bm n\otimes\bm n-\bm I).\label{eq:dd3bond}
\end{align}
Equation {\ref{eq:dd2bond}} and Equation {\ref{eq:dd3bond}} play a role in the derivation of bond-based thin plate model and bond-based gradient elasticity model in Section {\ref{sec:nonlocalplate}}. 

\subsection{NOM in 1D symmetric support}

To explicate the basic idea of NOM in symmetric support, we shall contemplate the NOM in one dimension and deduce the nonlocal bar/beam models. By virture of the Taylor series in one-dimensional space, the field difference $u_{ij}$ for bond $x_{ij}$ can be written as
\begin{align}
u_{ij}&\approx u'_ i x_{ij}+\frac 12 u_i'' x_{ij}^2,\label{eq:nom1d2}
\end{align}
where $u'_i$ and $u''_i$ denote the first-order and second-order derivatives, respectively, and the higher-order terms have been disregarded for the sake of simplicity. % in the derivation of Equation \ref{eq:nom1d2}. 
%For the case of one dimension, the gradient and curvature of a particle can be derived similarly as
%\begin{align}
%\frac{d u}{dx}&=\frac{1}{\int_{-\delta}^{\delta} \omega(x) x^2 dx}\int_{-\delta}^{\delta} \omega(x) u_{ij}x dx\notag\\
%\frac{d^2 u}{d x^2}&= \frac{1}{\int_{0}^{\delta} \omega(x) x^4 dx} \int_{0}^{\delta} \omega(x)(u_{ij}+ u_{ij'}) x^2 d x.\notag
%\end{align}
By computing the weighted first-order moment and second-order moment of Equation \ref{eq:nom1d2} in the interval $[x_i-\delta,x_i+\delta]$ and performing mathematical manipulation, one can derive the nonlocal first-order and second-order derivatives as
%\begin{subequations}
\begin{align}
u'_i&=\frac{\int_{x_i-\delta}^{x_i+\delta}\omega(r_{ij}) u_{ij} x_{ij} d x_j}{\int_{x_i-\delta}^{x_i+\delta} \omega(r_{ij}) x_{ij}^2 d x_j}.\notag\\
u_{i}''&=\frac{\int_{x_i-\delta}^{x_i+\delta}\omega(r_{ij}) u_{ij} x_{ij}^2 d x_j}{\int_{x_i}^{x_i+\delta} \omega(r_{ij}) x_{ij}^4 d x_j}=\frac{\int_{x_i}^{x_i+\delta}\omega(r_{ij})( u_{ij}+u_{ij'}) x_{ij}^2 d x_j}{\int_{x_i}^{x_i+\delta} \omega(r_{ij}) x_{ij}^4 d x_j}.\notag
\end{align}
%\end{subequations}
For each bond, the bond gradient and curvature are
\begin{align}
u_i'\langle x_{ij}\rangle=\frac{u_{ij} x_{ij}}{x_{ij}^2}=\frac{u_{ij}}{x_{ij}} %(\frac{x}{2\int_0^{\delta} \omega(x) x^2 dx}
,
u_i''\langle x_{ij}\rangle=\frac{( u_{ij}+u_{ij'}) x_{ij}^2}{x_{ij}^4}=\frac{ ( u_{ij}+u_{ij'})}{x_{ij}^2}.\label{eq:nom1ds}
\end{align}
\subsubsection{One-dimensional nonlocal bar}
Consider a one-dimensional nonlocal bar model with elastic modulus of $E$ and section area of $A$, we assume the bond energy density as
\begin{align}
\phi_{ij}=\frac 12 e_{ij} f_{ij} |r_{ij}|=\frac{1}{2}\omega(r_{ij}) c u_{ij}^2/|r_{ij}| ,\notag
\end{align}
where $e_{ij}=u_{ij}/|r_{ij}|$ is the relative strain and $ f_{ij}=\omega(r_{ij}) c e_{ij}$ is the bond force. The energy equivalence between local model and nonlocal model requires
\begin{align}
\int_{-\delta}^\delta \phi_{ij} dx=\frac 12 EA \varepsilon_{i}^2,\label{eq:ev1}
\end{align}
where $ \varepsilon_{i}$ is the local strain at point $x_i$.

In order to derive the specific form of bond force, a displacement field $u(x)=a x$ with constant gradient $u'(x)=a$ is assumed. Let $u_i=0, u_j=a x$, then $ u_{ij}=a x, e_{ij}= u_{ij}/|x|=a \mbox{ sign}(x), f_{ij}=\omega(x) c e_{ij}$. In 1D, only the elongation is involved.
The strain energy carried by a bond due to bond force and displacement becomes
\begin{align}
\phi_{ij}=\frac 12 e_{ij} f_{ij} |r_{ij}|=\frac 12 a^2 |x| c\omega(x).\notag
\end{align}
Here the process of doing work is considered, e.g. bond force $f_{ij}$ acting on distance $e_{ij} |r_{ij}|$. The energy equivalent in Equation \ref{eq:ev1} is calculated as
\begin{align}
\int_{-\delta}^\delta \frac 12 e_{ij} f_{ij} |x| dx=\int_{-\delta}^{\delta} \frac 12 a^2 |x| c \omega(x) dx=\frac 12 EA a^2 \to c=\frac{EA}{2 \int_{0}^{\delta} \omega(x) x dx}.\notag
\end{align}
The bond force in 1D is the variation of bond energy
\begin{align}
f_{ij}=\frac{\delta \phi_{ij}}{\delta u_{ij}} =\frac{ EA }{2\int_{0}^{\delta} \omega(x) x dx} \frac{\omega(r_{ij}) u_{ij}}{|r_{ij}|}.\label{eq:bbf1}
\end{align}

Another scheme to consider the bond energy is
\begin{align}
\phi_{ij}=\frac 12 e_{ij} f_{ij}=\frac 12 a^2 c \omega(x).\notag
\end{align}
The energy equivalence leads to 
\begin{align}
\int_{-\delta}^\delta \frac 12 e_{ij} f_{ij} dx=\int_{-\delta}^{\delta} \frac 12 a^2 c \omega(x) dx=\frac 12 EA a^2 \to c=\frac{EA}{2 \int_{0}^{\delta} \omega(x) dx}.\notag
\end{align}
The bond force of bond $ij$ is
\begin{align}
f_{ij}=\frac{\delta \phi_{ij}}{\delta u_{ij}}=\frac{EA}{2\int_{0}^{\delta} \omega(x) dx} \frac{ \omega(r_{ij}) u_{ij}}{ r_{ij}^2}.\label{eq:bbf2}
\end{align}
It is notable that the equations labeled as Equation \ref{eq:bbf1} and Equation \ref{eq:bbf2} exhibit equivalence under the condition that the weight function specified in Equation \ref{eq:bbf2} is assigned the form of $|r_{ij}|\omega(r_{ij})$. The force of direct bonding is applied to the material point denoted as $i$, while the force of reaction bonding is applied to the point denoted as $j$. The equivalence of bond $ij$ and bond $ji$ is observed, whereby the computation of bond $ji$ results in a reaction bond force of $-f_{ji}$ exerted on $i$, which conforms to the condition $-f_{ji}=f_{ij}$. Therefore, the governing equation for a nonlocal bar can be written as
\begin{align}
\int_{-\delta}^{\delta} 2 f_{ij} d x+b=\rho \ddot{u}_i,\notag
\end{align}
where $b$ and $\ddot{u}$ denote the body force and acceleration in 1D, respectively.

\subsubsection{One-dimensional nonlocal beam}
Consider a one-dimensional nonlocal beam model with elastic modulus denoted by $E$ and the second moment of area of the beam's cross section denoted by $I$, we can assume the bond bending energy density as
\begin{align}
\phi_{ij}=\frac 12\kappa_{ij} m_{ij}=\frac 12 \omega(r_{ij}) c \kappa_{ij}^2,\notag
\end{align}
where $\kappa_{ij}$ is the curvature of the bent bond and $ m_{ij}=\omega(r_{ij}) c \kappa_{ij}$ is the moment. The energy equivalence between local model and nonlocal model requires
\begin{align}
\int_{0}^\delta \frac 12 \kappa_{ij} m_{ij} dx=\frac 12 EI \kappa_{i}^2,\notag%\label{eq:ev2}
\end{align}
where $ \kappa_{i}$ is the local curvature at point $x_i$.

Let us assume a deflection field $u(x)=x^2/2$ with constant curvature $\kappa=u''(x)=1$. Let $x_i=0, x_j=x$, then $x_{ij}=x$, $ u_{ij}=x^2/2, \kappa_{ij}= ( u_{ij}+u_{ij'})/x^2=1, m_{ij}=\omega(x) c \kappa_{ij}= c\omega(x)$, $\frac 14\kappa_{ij} m_{ij}=\frac 14 c\omega(x)$.
The bending energy carried by a bond due to curvature is
\begin{align}
\frac 12 \kappa_{ij} m_{ij}=\frac 12 c\omega(x).\notag
\end{align}
The equivalent of bending energy in support to the local model can be simplified as
\begin{align}
\int_{0}^\delta \frac 12 \kappa_{ij} m_{ij} dx=\int_{0}^{\delta} \frac 12 c \omega(x) dx=\frac 12 EI\kappa^2=\frac 12 EI \to c=\frac{EI}{\int_{0}^{\delta} \omega(x) dx}.\notag
\end{align}

For a homogeneous beam with thickness $h$, the coefficient $c$ of different weight functions can be written as
%\begin{subequations}
\begin{align}
\frac{\omega(x) EI}{\int_{0}^{\delta} \omega(x) dx}=\begin{cases}
\frac{E h^3}{12 \delta} &\mbox{if } \omega(r)=1\\
\frac{E h^3 r}{24 \delta^2} &\mbox{if } \omega(r)=r.\\
\end{cases}\notag
\end{align}
%\end{subequations}
Formerly, the nonlocal curvature and moment can be explicitly written as
\begin{align}
\kappa_{ij}=\frac{( u_{ij}+u_{ij'})}{r_{ij}^2}, m_{ij}=\omega(r_{ij})\frac{( u_{ij}+u_{ij'})}{r_{ij}^2} \frac{EI}{\int_{0}^{\delta} \omega(x) dx}.\notag
\end{align}

The bent energy of bent bond is the multiplication of double volume $ \Delta x^2 $ and the bent energy density $\phi_{ij}$ as
\begin{align}
\phi_{ij}  \Delta x^2 =\frac 12 \kappa_{ij} m_{ij} \Delta x^2 =\frac 12 \frac{EI \omega(r_{ij}) }{\int_{0}^{\delta} \omega(x) dx} \frac{( u_{ij}+u_{ij'})^2}{r_{ij}^4} \Delta x^2,\notag
\end{align}
where $\Delta x$ is the volume of the material point.

The variation of $\phi_{ij} \Delta x^2$ reads

\begin{align}
\delta \phi_{ij}  \Delta x^2 &= \underbrace{\frac{EI \omega(r_{ij}) }{\int_{0}^{\delta} \omega(x) dx} \frac{( u_{ij}+u_{ij'})}{r_{ij}^4}\Delta x^2}_{\mbox{force due to bent: }f_{ijj'}} \cdot(\delta u_j+\delta u_{j'}-2 \delta u_i)\notag\\
&=f_{ijj'} \cdot \delta u_j+f_{ijj'} \cdot \delta u_{j'}-2 f_{ijj'} \cdot \delta u_i.\notag
\end{align}
Therefore, the bond forces adding to $i, j, j'$ due to bond curvature energy are $-2 f_{ijj'}, f_{ijj'},f_{ijj'}$, respectively.

In the context of utilizing an implicit algorithm, it is necessary to obtain the tangent stiffness matrix, which can be expressed through a second variation of $\phi_{ij}$:
\begin{align}
\delta^2 \phi_{ij}  \Delta x^2 &= \frac{EI \omega(r_{ij}) (\Delta x)^2 }{r_{ij}^4\int_{0}^{\delta} \omega(x) dx}(\delta u_j+\delta u_{j'}-2 \delta u_i)^2\notag\\
&=\begin{pmatrix}
\delta u_i\\
\delta u_j\\
\delta u_{j'}\\
\end{pmatrix}^T
\underbrace{(\frac{EI \omega(r_{ij}) (\Delta x)^2 }{r_{ij}^4\int_{0}^{\delta} \omega(x) dx})\begin{pmatrix}
4 & -2 &-2\\
-2& 1 & 1\\
-2 & 1 & 1\\
\end{pmatrix}}_{K_{ijj'}} \begin{pmatrix}
\delta u_i\\
\delta u_j\\
\delta u_{j'}\\
\end{pmatrix}.\notag
\end{align}
Herein, $K_{ijj'}$ denotes the tangent stiffness matrix of bent-bond $ijj'$.
\section{Nonlocal isotropic elasticity}\label{sec:nonisoe}
\subsection{Bond force in 3D}
%\begin{frame}{Nonlocal isotropic elasticity}
Consider the strain tensor projected on bond direction $\bm n_{ij}=(\cos \theta \sin \phi ,\sin \theta \sin \phi ,\cos \phi )$ in spherical polar coordinate based $\phi\in [0,\pi), \theta\in [0,2 \pi)$, the extension strain and shear strain along the bond direction are, respectively
\begin{subequations}
\begin{align}
\bm l_{ij}&=(\bm \varepsilon_i\cdot \bm n_{ij})\cdot (\bm n_{ij}\otimes \bm n_{ij})\\
\bm \gamma_{ij}&=(\bm \varepsilon_i\cdot \bm n_{ij})\cdot (\bm I-\bm n_{ij}\otimes \bm n_{ij}).
\end{align}
\end{subequations}
The relative strain vector $\bm \varepsilon_n=\bm \varepsilon_i\cdot \bm n_{ij}=\bm l_{ij}+\bm \gamma_{ij}$ and the relative displacement is $\bm u_{ij}=\bm \varepsilon_n r_{ij}=(\bm l_{ij}+\bm \gamma_{ij})r_{ij}$. 

\begin{figure}[htp]
\centering
\includegraphics[width=7cm]{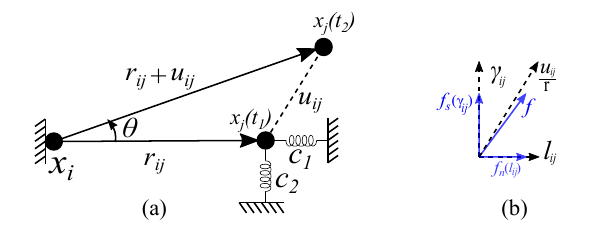}
\caption{Bond deformation with rotations, shear stiffness and extension stiffness.}
\label{fig:bondforces}
\end{figure}
As seen in Figure \ref{fig:bondforces}, {a local coordinate system can be established for any bond $\bm r_{ij}$. When a material point denoted by $\bm x_{j}(t_1)$ undergoes a relative displacement denoted by $\bm u_{ij}$, it transforms into a new position denoted by $\bm x_j(t_2)$. The bond's deformation is separated into directions that are parallel to and perpendicular to the bond direction. As illustrated in Figure} \ref{fig:bondforces}(a), the bond extension stiffness matrix and shear stiffness matrix are taken to be $c_1$ and $c_2$, respectively. The bond forces are taken into account for each direction as a function of the corresponding bond deformation as shown in Figure \ref{fig:bondforces}(b). To be more precise, we assume the bond force be the form
\begin{align}
\bm f_{ij}=\omega(r_{ij}) (c_1 \bm l_{ij}+c_2 \bm \gamma_{ij}).
\end{align}
%where $\omega(r)$ is a weight function.
The energy density associated with the deformation of a bond is
\begin{align}
w_{ij}=\frac 12\bm f_{ij}\cdot \bm u_{ij}=\omega(r_{ij}) (c_1 \bm l_{ij}+c_2 \bm \gamma_{ij})\cdot (\bm l_{ij}+\bm \gamma_{ij})r_{ij}=\frac 12\omega(r_{ij}) r_{ij}(c_1 \bm l_{ij}\cdot\bm l_{ij}+c_2 \bm \gamma_{ij}\cdot \bm \gamma_{ij}).\notag
\end{align}
Then the nonlocal strain energy density at a point in support domain equalizes to the local strain energy density
\begin{align}
&W=\int_{\mathcal S_i} w_{ij} d V_j=\int_{\mathcal S_i} \frac 12 r_{ij}\omega(r_{ij}) (c_1 \bm l_{ij}\cdot\bm l_{ij}+c_2 \bm \gamma_{ij}\cdot \bm \gamma_{ij}) dV_j\notag\\
&=W_{local}=\frac 12\bm \sigma:\bm \varepsilon=(\lambda \mbox{ Tr}(\bm \varepsilon)\bm I+2\mu \bm \varepsilon):\bm \varepsilon,\notag
\end{align}
where $\lambda,\mu$ are Lame constants.

For any $\bm \varepsilon$, using undetermined coefficient method yields

\begin{align}
c_1=\frac{ E}{\alpha (1 -2 \nu )},c_2=\frac{ E (1-4 \nu)}{ \alpha (\nu +1) (1-2 \nu)},
\end{align}
where $\alpha=\int_0^{\delta } \frac{4}{3}\pi r^3 \omega (r) dr$, and elastic modulus $E$ and Poisson's ratio $\nu$ are used to replace the Lame constants by $\lambda=\frac{E}{(1-2 \nu)(1+\nu)},\mu=\frac{E}{2(1+\nu)}$.

When the weight function $\omega(r_{ij})=1$, the coefficients become
\begin{align}
c_1=\frac{3 E}{\pi\delta ^4 (1 -2 \nu )},c_2=\frac{3 E (1-4 \nu)}{\pi \delta ^4 (\nu +1) (1-2 \nu)},
\end{align}
which are the same as the extended bond-based PD in \cite{zhu2017peridynamic}. The values of $c_1$ or $c_2$ are halved in the present study compared to those reported in \cite{zhu2017peridynamic}, due to the inclusion of direct bond force and reaction bond forces. Overall, the bond deformation and bond force, when taking into account the weight function, can be described as follows:
\begin{subequations}
\begin{align}
\bm l_{ij}&=\frac{\bm u_{ij}\cdot \bm n_{ij}}{r_{ij}} \bm n_{ij}\\
\bm \gamma_{ij}&=\frac{\bm u_{ij}}{r_{ij}}-\bm l_{ij}=\frac{\bm u_{ij}}{r_{ij}}-\frac{\bm u_{ij}\cdot \bm n_{ij}}{r_{ij}} \bm n_{ij}\\
\bm f_{ij}&=\omega(r_{ij}) (c_1 \bm l_{ij}+c_2 \bm \gamma_{ij}).\label{eq:fij3d}
\end{align}
\end{subequations}
And the corresponding governing equations are
\begin{align}
\int_{\mathcal S_i} 2 \bm f_{ij} d V_j+\bm b=\rho\ddot{\bm u}_i.
\end{align}

\subsection{Bond force in 2D}
For the case of plane stress condition, the material constitutive in local form is
\begin{align}
\bm \sigma=\frac{E}{1-\nu^2} (\nu \mbox{tr} \bm\epsilon \bm I_{2x2}+(1-\nu)\bm \varepsilon).\notag
\end{align}
The equivalence of strain energy density for arbitrary strain tensor leads to
\begin{align}
c_1=\frac{E}{ \alpha (1-\nu)}, c_2=\frac{E(1-3\nu)}{\alpha (1-\nu^2)},\label{eq:c1c2stress}
\end{align}
where $\alpha=\int_{0}^\delta \pi\omega(r) r^2 d r$.

Similarly, for plane strain condition, the material constitutive in local form is
\begin{align}
\bm \sigma=\frac{E}{(1+\nu)(1-2\nu)} (\nu \mbox{tr} \bm\epsilon \bm I_{2x2}+(1-2\nu)\bm \varepsilon).\notag
\end{align}
The energy equivalent gives the coefficients as
\begin{align}
c_1=\frac{E}{ \alpha (1-\nu-2 \nu^2)}, c_2=\frac{E(1-4\nu)}{\alpha (1-\nu-2\nu^2)},\label{eq:c1c2strain}
\end{align}
where $\alpha=\int_{0}^\delta\pi \omega(r) r^2 d r$. The bond-based governing equations for plane stress or plane strain can be expressed as Equation  \ref{eq:fij3d} with the utilization of coefficients derived from Equation  \ref{eq:c1c2stress} or Equation  \ref{eq:c1c2strain}.

The weighted bond-based nonlocal elasticity in 1D, 2D, and 3D is obtained by considering the local energy and assuming suitable bond deformation and bond force, while taking into account the energy equivalence. The energy associated with each bond is contingent solely upon its deformation, rendering it separable. However, the cumulative energy of all bonds restores isotropic elasticity.

\textbf{Remarks regarding the implementation}: In the preceding definition of bond force, it was established that each bond operates independently of the others, thereby minimizing interference and significantly enhancing numerical stability during bond breakage. The definition, however, is contingent upon the spherical support or horizon. For the purpose of facilitating numerical implementation, it is assumed that all particles possess an identical volume and support radius. The domain of interest is discretized through the utilization of uniform lattices. Particles at different locations exhibit identical coefficients for bonds that are oriented in the same direction and have the same distance.

%@article{Madenci2021Sep,
%	author = {Madenci, Erdogan and Barut, Atila and Phan, Nam},
%	title = {{Bond-Based Peridynamics with Stretch and Rotation Kinematics for Opening and Shearing Modes of Fracture}},
%	journal = {J. Peridyn. Nonlocal Model.},
%	volume = {3},
%	number = {3},
%	pages = {211--254},
%	year = {2021},
%	month = sep,
%	issn = {2522-8978},
%	publisher = {Springer International Publishing},
%	doi = {10.1007/s42102-020-00049-4}
%}
%
%\cite{Madenci2021Sep} broke the bond completely, so it cannot resist the compressive contact force. Or friction force.
\subsection{Two damage rules based on critical energy release rate}
In this subsection, by relating the critical shear strain or critical normal strain to the energy release rate, two damage rules are proposed.
\begin{figure}[htp]
\centering
\includegraphics[width=8cm]{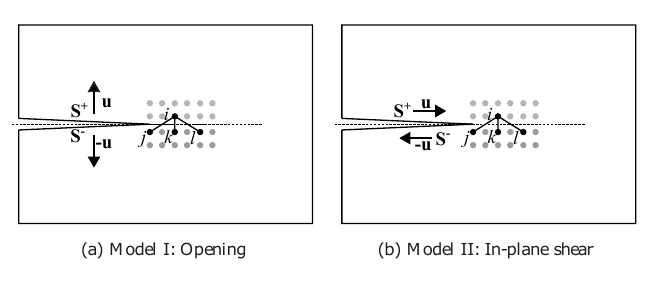}
\caption{Deformation of bond with shear deformation or tensile deformation, {where $S^+$ and $S^-$ denote the upper and lower crack surfaces, respectively.}}
\label{fig:tensileShearBond}
\end{figure}
\subsubsection{Critical normal strain damage rule}

Bond-based PD offers certain benefits, including the ability to maintain bond energy independence and a straightforward damage criterion that relies on critical stretch. Despite the significant perturbation caused by the direct neighbor cutting operation, the numerical stability of the system remains well preserved. However, in more general cases, there exist multiple impediments to the elimination of bonds. The deformation of individual bonds at the crack tip in the bond-based model that incorporates rotation for open-mode fracture is intricate. As illustrated in Figure \ref{fig:tensileShearBond}(a), the bond between nodes $i$ and $k$ is subjected to tensile deformation, while the bond between nodes $i$ and $j$ undergoes either shear deformation or a combination of shear and tensile deformation. The feasibility of implementing a criterion based on the stretch or rotation rule for severing the bond between nodes $i$ and $k$ is questionable, given that said bond is situated within an open-mode fracture. The coexistence of shear bonds and tensile bonds is apparent in the basic fracture mode. In the scenario depicted in Figure \ref{fig:tensileShearBond}(b), certain bonds exhibit compressive shear deformation, exacerbating the situation due to the sudden removal of internal bond force perturbation. The technique for bond removal relies heavily on geometric and intuitive procedures, yet it is deficient in a solid theoretical foundation. Several authors have investigated the shear and tensile deformation states in their research \cite{ni2019static,zhu2017peridynamic,zhang2019new,madenci2021bond}. By sorting these states and identifying the most vulnerable bonds, they have implemented a global iteration process to address the issue of damage. The complexity of these methods is relatively high due to their strong reliance on bond sequences, rendering them difficult to analyze theoretically. Furthermore, the construction of the global tangent stiffness matrix and its associated numerical methods incur significantly higher costs compared to the explicit time integration approach.

To facilitate automated fracture modeling, it is advisable to minimize the complexity of bond breakage. The phase field scheme proposed by Miehe et al. \cite{miehe2010thermodynamically} has been utilized to incorporate certain concepts into our work. The phase field model takes into account principal strains, which are not influenced by the shear strain in that particular direction. The numerical stability of the degradation of strain energy by the principal strain is commendable. In the context of bond-based nonlocal elasticity, it is typical for each bond to exhibit both axial and shear deformation. However, the accurate prediction of the magnitude of shear deformation, specifically the rotation, remains a challenge due to the potential impact of rigid rotation. In terms of discretization, the bonds form a loop in all bond directions in support. The computation of eigenvalue decomposition for strain tensors is not a mandatory requirement. In cases where the bond direction aligns with the principal strain direction, the scenario bears resemblance to the phase field model. In contrast to the bond-based model that incorporates stretch resulting from shear deformation and normal deformation, as well as the critical rotation model proposed in previous literature \cite{zhang2019new,madenci2021bond}, our approach solely considers bond-directional strain while disregarding shear deformation. The alteration in strain energy in this context exhibits a resemblance to the phase field model, as it solely accounts for the energy along the principal strain direction.

For each bond direction, the interaction status is determined through a bond status parameter given by
\begin{align}
\mu(\bm x_j-\bm x_i)=\begin{cases} 1, s_n(t) <s_n^c\\
0, \underset{0\leq t \leq T}{\max} s_n(t) \geq s_n^c,
\end{cases}\notag
\end{align}
where $s_n(t)$ is the strain at time $t$ along the initial bond direction and $s_n^c$ is the critical bond stretch determined by the Griffith energy release rate.
The local damage is evaluated as
\begin{align}
\phi(\bm x_i)=1-\frac{\int_{\mathcal S_i} \mu(\bm x_j-\bm x_i) d V_j}{\int_{\mathcal S_i} d V_j}.\notag
\end{align}

When considering the deformation along the principal strain direction, the deformation can be simplified into 1D with cross-section area $A$.
Consider the deformation in 1D, in order to form a crack surface, half support should be cut. The equivalence of fracture energy and strain energy in half support is
\begin{align}
\frac 12 G_c A=\frac 12 K \varepsilon^2 A \delta \to \varepsilon=\sqrt{\frac{G_c}{K \delta}}, \label{eq:GI}
\end{align}
where $K=\frac{E}{3(1-2\nu)} $ is the bulk modulus of the material, {$\delta$ is the radius of support $\mathcal S$ and $G_c$ denotes the critical energy release rate}. Therefore, we select the critical normal strain as
\begin{align}
s_n^c=\sqrt{\frac{G_c}{K \delta}}=\sqrt{\frac{3(1-2\nu)G_c}{E \delta }}.
\end{align}
The rule that stems from the one-dimensional scenario appears to be straightforward; however, it can yield nearly equivalent precision outcomes as the phase field approach through finite element techniques under certain circumstances, as evidenced by the numerical illustrations.

\subsubsection{Critical shear strain damage rule}
The maximal shear strain direction is another direction that is akin to the principal strain direction. Shear fractures are often caused by shear strain in various materials. When examining deformation in one dimension, it is natural to note that the fracture energy and strain energy are equivalent in half support
\begin{align}
\frac 12 G_{II} A=\frac 12 \mu \varepsilon_s^2 A \delta \to \varepsilon_s=\sqrt{\frac{G_{II}}{\mu \delta}}, \label{eq:GII}
\end{align}
where $G_{II}$ is the critical energy release rate for mode II fracture and $\mu$ the shear modulus. Therefore, the critical shear strain is selected as
\begin{align}
s_t^c=\sqrt{\frac{G_{II}}{\mu \delta}}=\sqrt{\frac{2(1+\nu)G_{II}}{E \delta }}.\label{eq:secrit}
\end{align}

\subsection{Plasticity for bond element}
{The incorporation of plasticity deformation in the bond-based nonlocal model is feasible by specifying the appropriate yielding function and flow rule based on the plasticity theory} \cite{hill1998mathematical,liu2020ordinary,cervera2022mechanics}. 
  
Let $(n,m,t)$ be a set of orthogonal local axes, with $\bm e_n,\bm e_m,\bm e_t$ being the normal vector, the shear-direction and out-of-plane direction of the bond element, respectively.
The kinematics of a bond element is
\begin{align}
\bme_n=l \bm e_n+\gamma \bm e_m,\notag
\end{align}
where $l=\frac{\bm u_{ij}}{r}\cdot \bm e_n, \gamma=\frac{\bm u_{ij}}{r}\cdot \bm e_m$.

In a local coordinate system, the strain and force are
\begin{align}
\bme=(l, \gamma)^T, \bms=(\sigma,\tau)^T=\underbrace{\begin{pmatrix}
c_1 & 0\\
0 & c_2
\end{pmatrix}}_{\mathbb E_0} \begin{pmatrix}
l\\
\gamma
\end{pmatrix},\notag
\end{align}
where the vector $\bms,\bme$ represent the force vector and strain vector; $\mathbb E_0$ is the second-order material tensor in bond local coordinates.

For elastoplastic models, the constitutive relation of a bond element in local coordinate can be expressed in rate form as
\begin{align}
\dbme=\dbme^e+\dbme^p,\dbms=\mathbb E_0 \cdot (\dbme-\dbme^p),\notag
\end{align}
where $\bme^e$ and $\bme^p$ being the elastic and plastic parts of the strain tensor.

Without loss of generality, the plastic strain rate is given by the following flow rule based on the plastic potential function $f^p(\bms,q)$
\begin{align}
\dbme^p=\dot{\lambda}\underbrace{\pfrac{f^p}{\bms}}_{\Lambda^p}, \dot{\kappa}=-\dot{\lambda} \pfrac{f^p}{q},\notag
\end{align}
for the plastic multiplier $\dot{\lambda}$ satisfying the classical Karush-Kuhn-Tucker conditions
\begin{align}
\dot{\lambda}\geq 0, f(\bms,q)\leq 0,\dot{\lambda} f(\bms,q)=0,\notag
\end{align}
where a force-based yield function $f(\bms,q)\leq 0$, with $q$ being the force-like internal variable (yield force) conjugate to the strain-like one $\kappa$ which measures the plastic state; $\Lambda^p:=\pfrac{f^p}{\bms}$ is the plastic flow direction. For associated plasticity, the potential function $f^p(\bms,q)$ is identical or proportional to the yield function $f(\bms,q)$.

Then the force state rate can be written as
\begin{align}
\dbms=\mathbb E_0 \cdot (\dbme-\dot{\lambda} \Lambda^p ).\notag
\end{align}

Plastic yielding occurs when the yield condition $f(\bms,q)=0$ is activated, i.e. $\dot{\lambda}>0$. Follow from the consistency condition $\dot{f}=0$ gives
\begin{align}
\dot{\lambda}=\frac{\Lambda\cdot \mathbb E_0 \cdot \dbme}{\Lambda\cdot \mathbb E_0 \cdot \Lambda^p+h H h^p},\notag
\end{align}
for the derivative $\Lambda:=\pfrac{f}{\bms}$ and $h=-\pfrac{f}{q}$ of yield function $f(\bms,q)$ and hardening/softening modulus $H:=\pfrac{q}{\kappa}$.

The corresponding constitutive relation in rate form then reads
\begin{align}
\dbms=\mathbb E\cdot (\dbme-\dbme^p)=\mathbb E^{ep}\cdot \dbme,\notag
\end{align}
where the second-order elastoplasticity tangent $\mathbb E^{ep}$ is expressed as
\begin{align}
\mathbb E^{ep}=\mathbb E_0-\frac{(\mathbb E_0\cdot \Lambda^p)\otimes ( \Lambda\cdot \mathbb E_0)}{\Lambda\cdot \mathbb E_0 \cdot \Lambda^p+h H h^p}.\notag
\end{align}

In anlogy with the Mohr–Coulomb yield and plasticity, the yield function and plastic potential function are assumed to have the form
\begin{align}
f^p=f(\sigma,\tau,q)=a \sigma +b \sqrt{\tau^2}-q(\kappa),\notag
\end{align}
{where $a,b$ are material coefficients for the plasticity on a bond.} 
With this yield function, the flowing direction $\Lambda$ and elastoplasticity tangent can be explicitly written as
\begin{align}
\Lambda=(a, b \mbox{ sign}(\tau)),\notag
\end{align}
\begin{align}
\mathbb E^{ep}=\omega(r)\Big(\begin{pmatrix}
c_1& 0\\
0 & c_2
\end{pmatrix}-\frac{1}{a^2 c_1+b^2 c_2}\begin{pmatrix}
a^2 c_1^2 & a b c_1 c_2 \mbox{ sign}(\tau)\\
a b c_1 c_2 \mbox{ sign}(\tau) & b^2 c_2^2
\end{pmatrix}\Big),\notag
\end{align}
where $\mbox{sign}(x)$ is the sign function.

\section{Higher-order nonlocal bond-based models}\label{sec:nonlocalplate}
The bond-based nonlocal model is not restricted in first-order. By making use of the bent-bond, the bond-based plate model and bond-based gradient elastic model will be derived in the following.
\subsection{Nonlocal isotropic thin plate}
%\begin{frame}{Nonlocal isotropic thin plate}
For deflection field $w(x,y)=\frac 12(a x^2+b y^2+2 c x y)$, where $(a,b,c)$ are arbitrary real numbers
, the second-gradient of deflection field, $\bm \kappa=\nabla\nabla w$, can be written as
\begin{align}
\bm \kappa=\begin{pmatrix}
\kappa_{11}& \kappa_{12} \\
\kappa_{12}& \kappa_{22} \\
\end{pmatrix}=\begin{pmatrix}
a& c \\
c& b \\
\end{pmatrix}\notag
\end{align}

Along with bond direction $\bm n=(\cos \theta ,\sin \theta )$, the orthogonal decomposition of bond curvature tensor given by Equation \ref{eq:dd2bond} is
\begin{align}
\bm \kappa_n&=(w_{ij}+w_{ij'})/r_{ij}^2 (4 \bm n\otimes \bm n-\bm I)\notag\\
&=\underbrace{3 (w_{ij}+w_{ij'})/r_{ij}^2 (\bm n\otimes \bm n)}_{\bm \kappa_{nn}}+\underbrace{(-(w_{ij}+w_{ij'})/r_{ij}^2)(\bm I-\bm n\otimes \bm n)}_{\bm \kappa_{ns}}.
\end{align}
%\begin{figure}[htp]
%\centering
%\includegraphics[width=8cm]{bondwithrotationgradient.pdf}
%\caption{Bond deformation with rotations, shear curvature stiffness and extension curvature stiffness.}
%\label{fig:bondforcesgrad}
%\end{figure}
In analogy to the normal-shear decomposition of the deformation in Section \ref{sec:nonisoe}, the bending moment for single bent bond can be assumed as
\begin{align}
\bm M_n=\omega(r_{ij})(c_1 \bm \kappa_{nn}+c_2 \bm \kappa_{ns}),
\end{align}
where $c_1,c_2$ are the material parameters to be determined. 
%\end{frame}
%\begin{frame}
The total energy carried by a point
\begin{align}
W_{nonlocal}&=\int_{\mathcal S^+} \frac 12 \bm M_n: \bm \kappa_n dV\notag\\
&=\int_{\mathcal S^+} \frac 12 \omega(r) (w_{ij}+w_{ij'})^2/r_{ij}^4 (9 c_1+c_2) dV.\notag
\end{align}
$W_{nonlocal}=W_{local}:=\frac 12\bm M:\bm \kappa$ for any field yields $(9 c_1+c_2)=\frac{16 D_0}{3 \alpha}, \nu=1/3$, where $\alpha=\int_0^\delta \omega(r)\pi r\, d r$.
Therefore, the equivalent curvature and moment for a bond are
\begin{subequations}
\begin{align}
\kappa_{ij}&=(w_{ij}+w_{ij'})/r_{ij}^2\\
m_{ij}&=\omega(r)\frac{16 D_0}{3 \alpha} (w_{ij}+w_{ij'})/r^2=\omega(r)\frac{E t^3}{2 \alpha} (w_{ij}+w_{ij'})/r_{ij}^2,
\end{align}
\end{subequations}
where $D_0=\frac{E t^3}{12(1-\nu^2)}=\frac{3}{32} E t^3$ and $t$ is the thickness of the plate. This is the bond-based version of nonlocal thin plate. Only the Poisson's ratio of 1/3 can be modeled.

The corresponding bond force can be derived by considering the first variation of the bond energy
\begin{align}
f_{ijj'}=\omega(r_{ij})\frac{E t^3}{2 \alpha} \frac{(w_{ij}+w_{ij'})}{r_{ij}^4}.
\end{align}

\subsubsection{Cohesive damage model for bent bond}
\begin{figure}[htp]
\centering
\includegraphics[width=7cm]{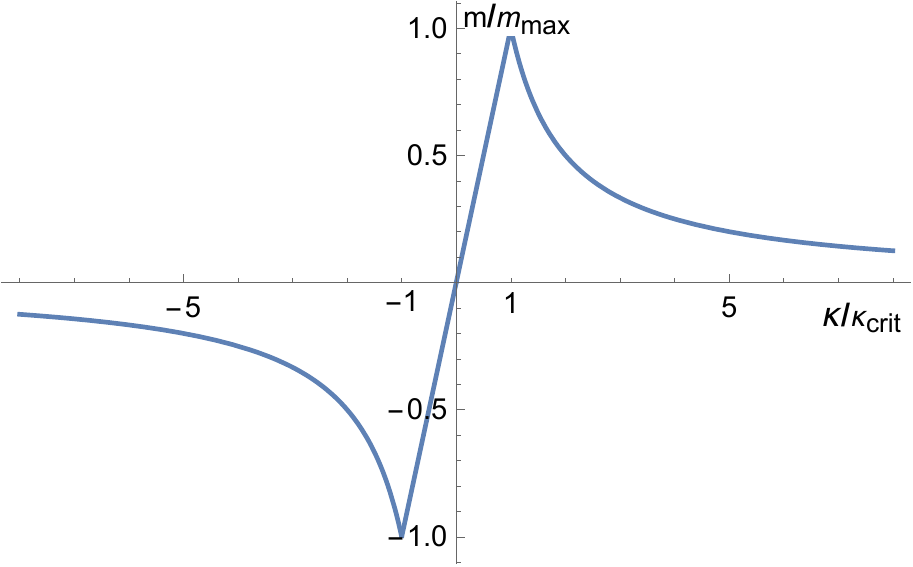}
\caption{Moment and curvature relation in a cohesive model with $\kappa_{crit}$ and $m_{max}$ being critical curvature and critical moment.}
\label{fig:mkCohesive}
\end{figure}

In order to introduce the localization, the moment is calculated as
\begin{align}
m_{ij}=\omega(r)\frac{E t^3}{2 \alpha} \mbox{sign}(\kappa_{ij})\min(|\kappa_{ij}|, \frac{\kappa_{crit}^2}{|\kappa_{ij}|}),\label{eq:mkCohesive}
\end{align}
where $\kappa_{crit}$ is the critical curvature when softening of force occurs. The curve of Equation \ref{eq:mkCohesive} is plotted in Figure \ref{fig:mkCohesive}.

\subsection{Nonlocal isotropic gradient elasticity}

Similar to the nonlocal thin plate, we consider one field in gradient elasticity. By virtue of spherical coordinate system in 3D, the bond can be represented by coordinates $(r,\theta,\phi)$. Along the bond-direction, a local coordinate system can be created with orthogonal unit basis vectors $\bm n_1, \bm n_2,\bm n_3$ as
\begin{align}
\bm n_1=(\cos\theta \sin\phi,\sin\theta\sin\phi,\cos\phi),\notag\\
\bm n_2=(\cos\theta \cos\phi,\cos\phi\sin\theta,-\sin\phi),\bm n_3=(-\sin\theta,\cos\theta,0).\notag
\end{align} 
For any field $u(x,y,z)=\frac{1}{2} (a x^2+b y^2+c z^2+2 d x y+2 f x z+2 g y z)$ in $x$-direction, the curvature tensor is $\bm \kappa=((a,d,f),(d,b,g),(f,g,c))$, where $(a,b,c,d,f,g)$ are arbitrary real numbers.

The orthogonal decomposition of nonlocal Hessian on a bond given by Equation \ref{eq:dd3bond} is
\begin{align}
\bm \kappa_n&=( u_{ij}+u_{ij'})/r_{ij}^2(5\bm n_1\otimes \bm n_1-\bm I)\notag\\
&=\underbrace{4 ( u_{ij}+u_{ij'})/r_{ij}^2 \bm n_1\otimes \bm n_1}_{\bm \kappa_1}+\underbrace{(-( u_{ij}+u_{ij'})/r_{ij}^2) \bm n_2\otimes \bm n_2}_{\bm \kappa_2}\notag\\
&+\underbrace{(-( u_{ij}+u_{ij'})/r_{ij}^2) \bm n_3\otimes \bm n_3}_{\bm \kappa_3}.\notag
\end{align}

The bending moment of the bond is assumed as
\begin{align}
\bm M_n&=\omega(r_{ij})\Big(\underbrace{c_1 4 ( u_{ij}+u_{ij'})/r_{ij}^2 \bm n_1\otimes \bm n_1}_{\bm M_1}\notag\\
&+\underbrace{c_2 (-( u_{ij}+u_{ij'})/r_{ij}^2) \bm n_2\otimes \bm n_2}_{\bm M_2}+\underbrace{c_2(-( u_{ij}+u_{ij'})/r_{ij}^2) \bm n_3\otimes \bm n_3}_{\bm M_3}\Big),\notag
\end{align}
where $c_1,c_2$ are the unknown curvature stiffness and here we assume the stiffnesses in $\bm M_2$ and $\bm M_3$ are the same.

%\end{frame}
%\begin{frame}
The bent energy carried by a bond is
\begin{align}
w_n=\frac 12 \bm M_n:\bm \kappa_n =\omega(r_{ij})(8 c_1 + c_2)( u_{ij}+u_{ij'})^2/r_{ij}^4.
\end{align}
The bent energy carried by a point is the summation of all bent bonds:
\begin{align}
W=\int_{\mathcal S^+} w_n dV=\frac{1}{15} \alpha (8 {c_1}+{c_2}) (3 a^2+2 a (b+c)+3\notag\\
b^2+2 b c+3 c^2+4 (d^2+f^2+g^2)),\notag
\end{align}
where $\alpha=\int_0^{\delta } \pi r^2 \omega(r) dr$.

For the given material constitution of gradient strain energy as $\bm M=\ell^2(\lambda \mbox{Tr}(\bm \kappa)\bm I+2 \mu \bm \kappa)$, where $\ell$ is length scale parameter in gradient elasticity, the gradient strain energy in local theory for any curvature deformation can be simplified as 
\begin{align}
W_{local}=\frac 12 \bm M:\bm \kappa=\frac{1}{2} \ell^2 \Big(a^2 (\lambda +2 \mu )+2 a \lambda (b+c)\notag\\
+b^2 (\lambda +2 \mu )+2 b c \lambda +c^2 \lambda +2 c^2 \mu +4 d^2 \mu +4 f^2 \mu +4 g^2 \mu \Big).\notag
\end{align}
%\end{frame}
%\begin{frame}

The energy equivalence $W=W_{local}$ for any $\bm \kappa$ leads to
\begin{align}
(8c_1+c_2)=\frac{15 \ell^2 \mu}{2 \alpha},\lambda=\mu \to \nu=\frac 14.\notag
\end{align}

Therefore, the curvature and bending moment of $u$ field for bond $ij$ are, respectively, 
\begin{align}
\kappa^u_{ij}=( u_{ij}+u_{ij'})/r_{ij}^2, m_{ij}^u=\omega(r_{ij})\frac{15 \ell^2 \mu}{\alpha} ( u_{ij}+u_{ij'})/r_{ij}^2.\notag
\end{align}
And the bent-bond energy becomes
\begin{align}
\phi_{ij}=\frac 12 \omega(r_{ij})\frac{15 \ell^2 \mu}{\alpha} ( u_{ij}+u_{ij'})^2/r_{ij}^4.\notag
\end{align}

Hence, the corresponding bond force is
\begin{align}
f_{ijj'}^u=\frac{\partial \phi_{ij}}{\partial (u_{ij}+u_{ij'})}= \omega(r_{ij})\frac{15 \ell^2 \mu}{\alpha} ( u_{ij}+u_{ij'})/r_{ij}^4.\notag
\end{align}
The curvature bond force follows the direction of $u$ field.

For field $v,w$, the same conclusions can be obtained. In sum, the curvature bond force of bond-based gradient elasticity is
\begin{align}
\bm f_{ijj'}=\omega(r_{ij})\frac{15 \ell^2 \mu}{\alpha} (\bm u_{ij}+\bm u_{ij'})/r_{ij}^4.
\end{align}

The governing equations of bond-based gradient elasticity become
\begin{align}
\underbrace{\int_{\mathcal S_i} 2\bm f_{ij} dV_j}_{first-order\,contribution}+\underbrace{\int_{\mathcal S_i^+} 2\bm f_{ijj'}d V_j-\int_{\mathcal S_i}\bm f_{jii'}d V_j}_{second-order\, contribution}+\bm b=\rho \ddot{\bm u}_i.
\end{align}
The first-order contribution contains the conventional bond forces and the second-order contribution is the bent bond force due to curvatures. It is noteworthy to mention that the force $\bm f_{jii'}$ is regarded as a bent bond force defined by $j,i$ and $i'$, where $i\in \mathcal S_j$ and $i'\in \mathcal S_j$. Additionally, the point $i'$ is the mirrored image of point $i$ with respect to $j$, e.g. $\bm r_{ji'}=-\bm r_{ji}$.

\section{Numerical examples}
The Verlet-velocity explicit time integration algorithm is utilized to conduct numerical examples. In certain scenarios, the quasi-static state can be attained through the gradual imposition of velocity boundary conditions. The summation of internal forces of the designated particle set is performed to obtain the reaction forces, which are subsequently subjected to the boundary conditions. 
\subsection{Simply supported beam}
The bent bond, which is characterized by three points, exhibits first-order derivative immunity under full support. Nonetheless, in the case of material points located in close proximity to the boundaries, the support domain becomes incomplete. Supplementary particles are incorporated beyond the borders of the support domain to ensure its completeness. The condition of simply supported boundary is satisfied by
\begin{align}
w(0)=w(L)=0.\notag
\end{align}
The function of additional particles is to make sure the half support $\mathcal S^+$ is well defined.
The full implementation code of the simply supported beam can be found by the link \url{https://github.com/hl-ren/Nonlocal_beam}.

\begin{figure}[htp]
\centering
\includegraphics[width=8cm]{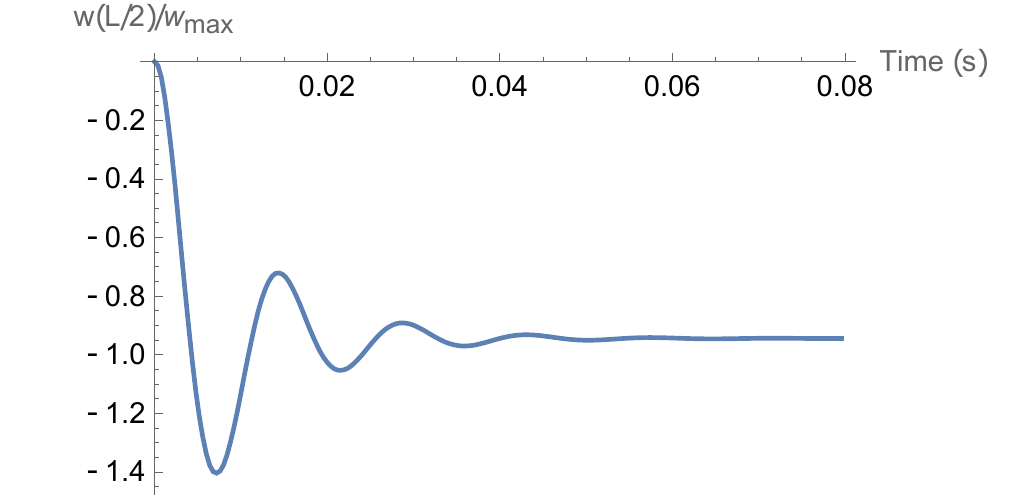}
\caption{The evolution of deflection of midpoint .}
\label{fig:bbbeam25n2mid}
\end{figure}

\begin{figure}[htp]
\centering
\includegraphics[width=10cm]{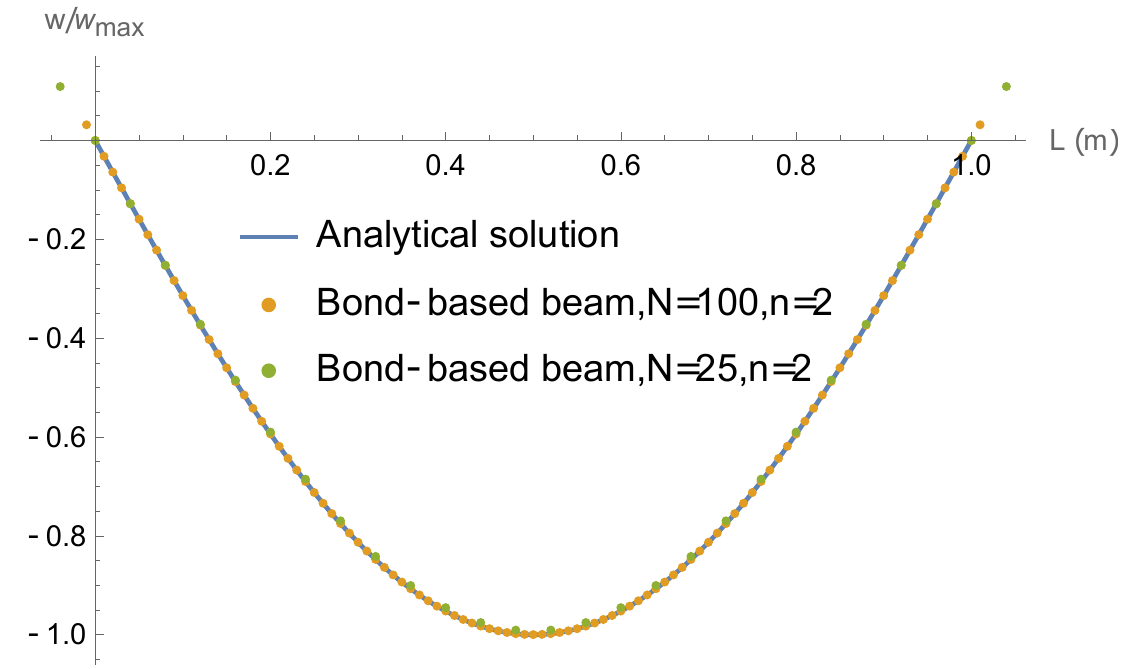}
\caption{The comparison of exact solution and bond based beam $N=100$ and $N=25$ .}
\label{fig:bbbeam25-100n2}
\end{figure}

The present example involves material parameters of $E=30\times 10^9$ Pa, a beam length of $L=1$, and a thickness of $h=0.05$. A damping mechanism with a coefficient of $-300\dot{w}$ is employed to achieve convergence of the dynamic solution to the static outcome. The plot in Figure \ref{fig:bbbeam25n2mid} illustrates the evolution of deflection at the midpoint with the presence of damping. The graphical representation of the ultimate displacement of the beam can be observed in Figure \ref{fig:bbbeam25-100n2}. The results indicate that the computed solution for a system comprising of $N=25$ material points and $\delta =2\Delta x$ exhibits a high degree of proximity to the exact solution, where $\Delta x$ is the grid space in discretization.. The figure labeled as Figure \ref{fig:bbbeam25-100n2} demonstrates that the deflection of a beam with a discretization of $N=100$ is in good agreement with the exact solution. 

\begin{figure}[htp]
\centering
\includegraphics[width=7cm]{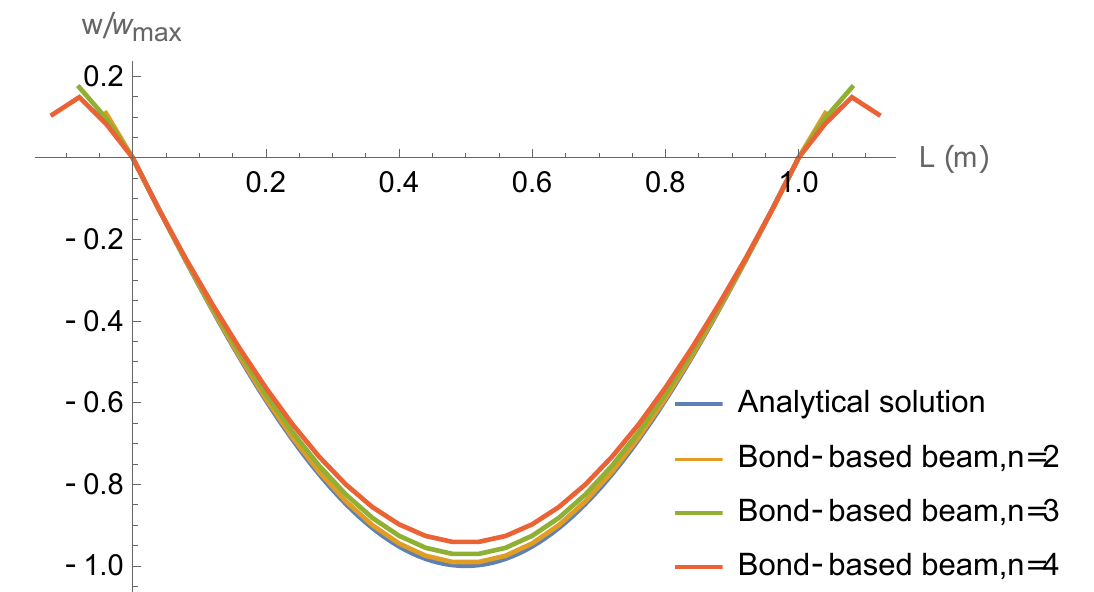}
\caption{The influence of support size in a bond based beam .}
\label{fig:bbbeam25n234}
\end{figure}

\begin{figure}[htp]
\centering
\includegraphics[width=7cm]{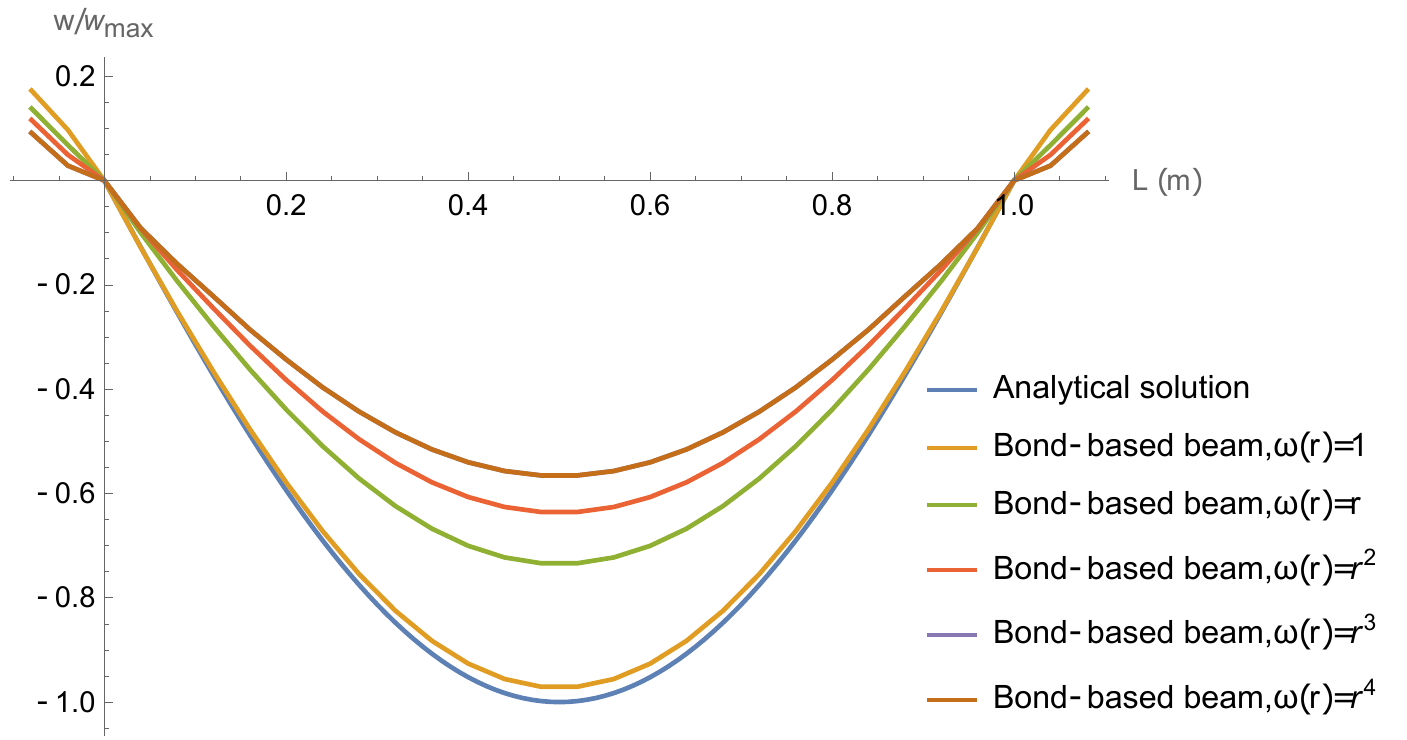}
\caption{The influence of weight function in a bond based beam.}
\label{fig:winfbb25}
\end{figure}

Figure \ref{fig:bbbeam25n234} examines the impact of support size. As the support size is increased, the beam experiences a marginal increase in stiffness. Figure \ref{fig:winfbb25} displays the impact of the weight function $\omega(r)=r^n$, where $n$ takes on values in the set $\{0,1,2,3,4\}$. The support size has been chosen as $\delta=3\Delta x$. It is evident that the weight function plays a crucial role in determining the deflection in this scenario. %The L2 norm of error for this agreement is $3.01\times 10^{-4}$.

\begin{figure}[htp]
\centering
\includegraphics[width=7cm]{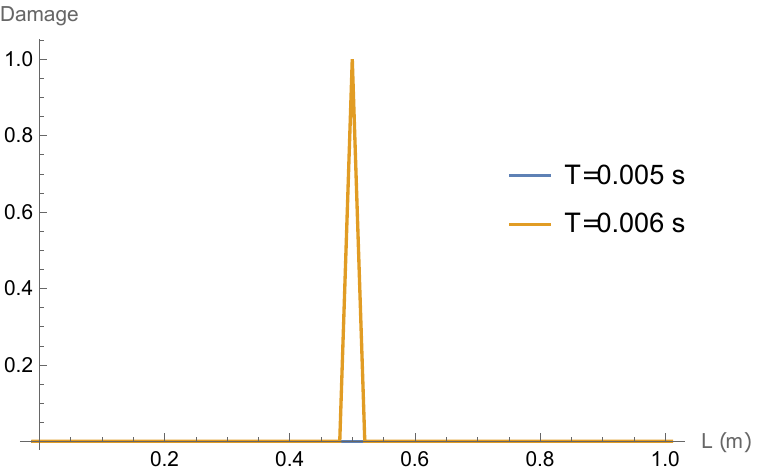}
\caption{Damage in bond based beam.}
\label{fig:bbbeam100damage}
\end{figure}

\begin{figure}[htp]
\centering
\includegraphics[width=7cm]{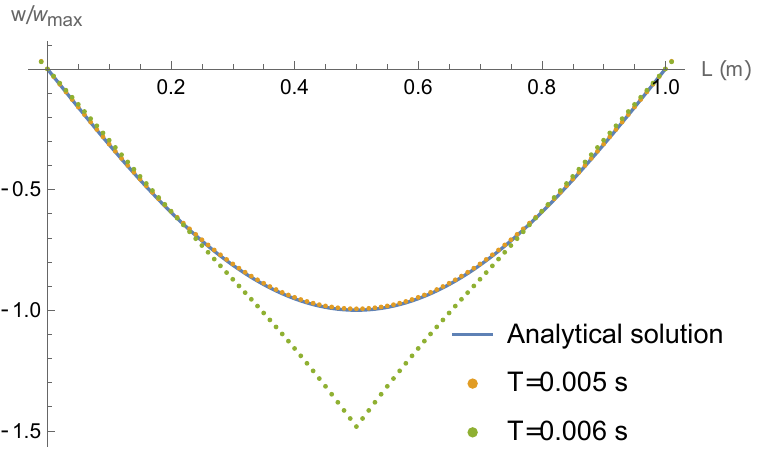}
\caption{Deflection of bond based beam.}
\label{fig:bbbeam100damageW}
\end{figure}

In order to model fracture in a thin beam, we applied the cohesive damage rule to model the fracture. We select the critical curvature tensor as $\kappa_{crit}=4\times 10^{-4}$ and use the damping coefficient $p=300$ for reducing oscillation. The damage distribution and displacement field at the $t=0.03$ seconds are shown in Figure \ref{fig:bbbeam100damage} and Figure \ref{fig:bbbeam100damageW}, respectively. It can be observed that the damage happens at the center of the beam.

\subsection{Single-edge-notched tension test}
\begin{figure}
	\centering
		\includegraphics[width=5cm]{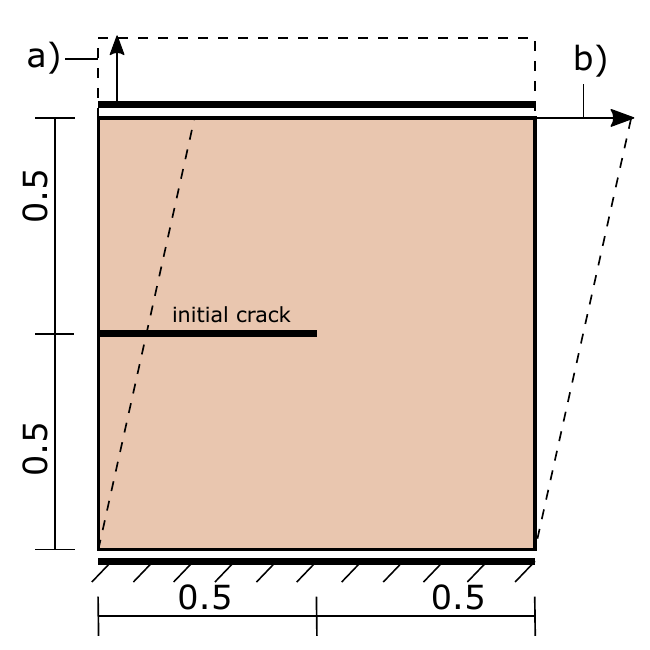}
	\caption{Single-edge-notched test. Geometry and Case a for tensile boundary condition and Case b for shear boundary condition.}
	\label{fig:notchPlate-4}
\end{figure}

In this subsection, we model the single-edge-notched tension test, which is a squared plate with initial notched crack as shown in Figure \ref{fig:notchPlate-4}. The material parameters are set as $\lambda=121.1538$ kN/mm$^2$ and $\mu=80.7692$ kN/mm$^2$ for elastic constants, $G_c=2.7\times 10^{-3}$ kN/mm for the critical energy release rate. These parameters are identical to that used in the small strain brittle fracture phase field in Ref \cite{miehe2010thermodynamically}. Two displacement conditions are tested: Case a) for tensile boundary condition and Case b) for shear boundary conditions. The plate is discretized with three settings: $60 \times 60$, $120\times 120$ and $200\times 200$ material points. The displacement load is monotonic applied with velocity boundary condition defined by
\begin{align}
v(t)=\begin{cases}
\frac{t}{t_0} v_{max} &\mbox{ if } t< t_0\\
v_{max} &\mbox{ otherwise}
\end{cases}\notag
\end{align}
with $t_0=1.0 \times 10^{-5} s$ and $v_{max}=2$ m/s.

In the case of tensile load, three discretizations are employed.
\begin{figure}
	\centering
		\includegraphics[width=12cm]{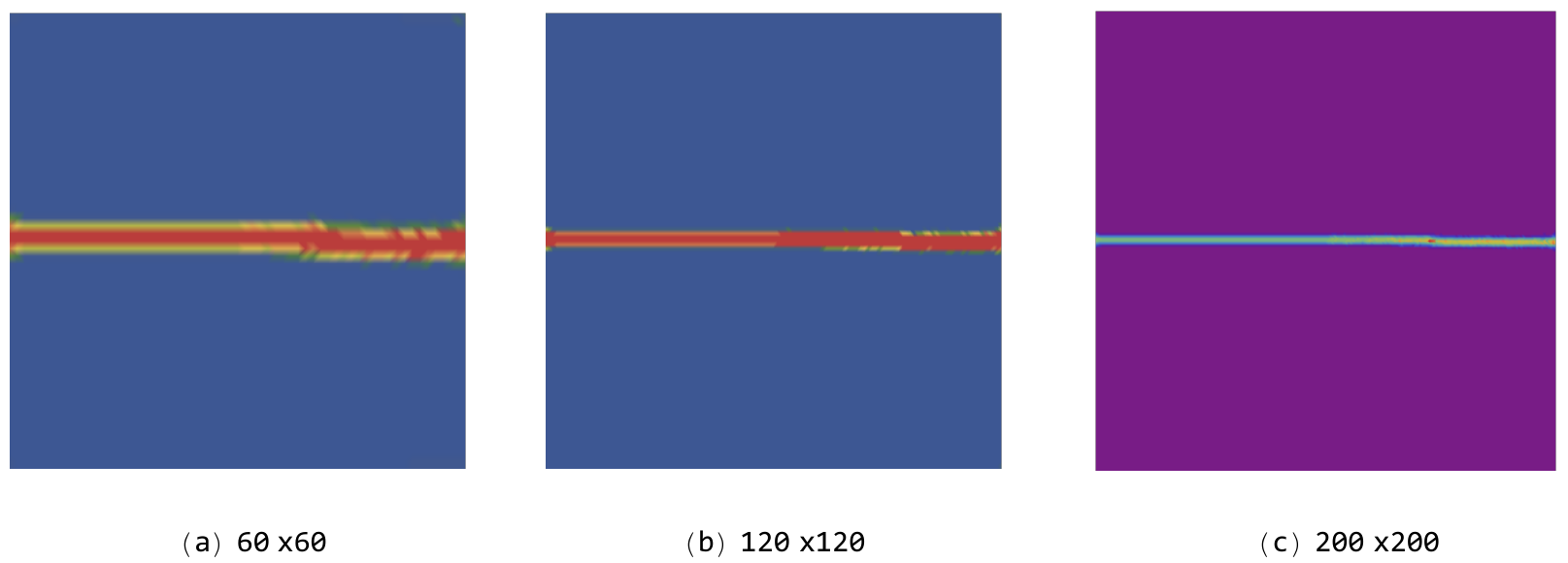}
	\caption{Single-edge-notched tension test, damage patterns for 60x60 particles, 120x120 particles and 200x200 particles.}
	\label{fig:tensileLoadDamage60-120}
\end{figure}
The damage patterns for particle discretization of 60x60, 120x120 and 200x200 are illustrated in Figure \ref{fig:tensileLoadDamage60-120}. The observed crack patterns exhibit a high degree of concurrence with those predicted by phase field approaches, and a more refined discretization can yield a sharper crack trajectory.
\begin{figure}
	\centering
		\includegraphics[width=14cm]{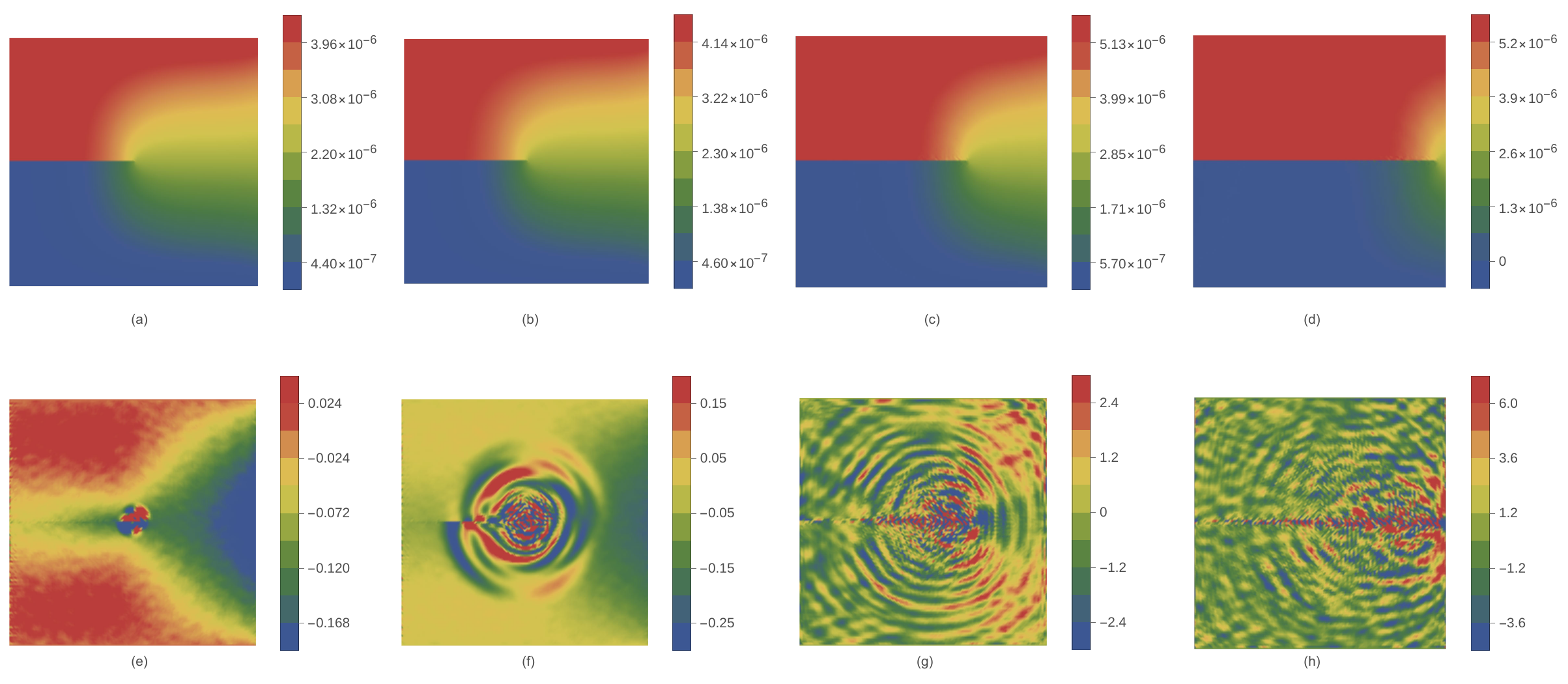}
	\caption{Single-edge-notched tension test: first row (a,b,c,d) denotes displacement field in $y$-direction and the second row (e,f,g,h) velocity in $x$-direction. Two figures of the same column correspond to the same time.}
	\label{fig:tensile120uyVx3}
\end{figure}
Figure \ref{fig:tensile120uyVx3} displays the temporal variations of the displacement field and velocity field. The initiation of the crack occurs at the point where the boundary displacement attains a value of $u_y=3.96 \times 10^{-3}$ mm, as evidenced by the anomalous velocity field surrounding the crack tip depicted in Figure \ref{fig:tensile120uyVx3}(e). During the stage of stable crack propagation, the velocity wave caused by cutting bond is prominently visible in Figure \ref{fig:tensile120uyVx3}(f). The presence of fractures significantly disrupts the velocity field, while the displacement field remains stable.

\begin{figure}
	\centering
		\includegraphics[width=8cm]{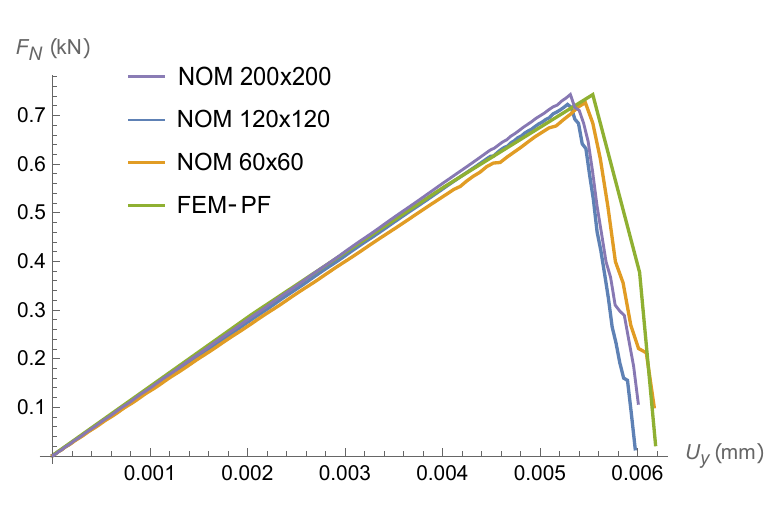}
	\caption{Single-edge-notched tension test, load curves for NOM and phase field by FEM.}
	\label{fig:tensileLoadCurve}
\end{figure}
Figure \ref{fig:tensileLoadCurve} displays the load curves pertaining to the tensile boundary. The criterion of maximal normal strain is obtained through a straightforward process, yet it has demonstrated remarkable efficacy in practical implementation. The tensile boundary condition test yielded load-curve results that were highly consistent with those obtained from the finite element phase field model. {The research conducted on three discretization settings indicates that the damage model exhibits good robustness towards the discretization employed.} The observation that the fracture model employing explicit time integration in the absence of damping exhibits a high degree of agreement with the phase field approach in the stationary scenario is noteworthy.

\begin{figure}
	\centering
		\includegraphics[width=12cm]{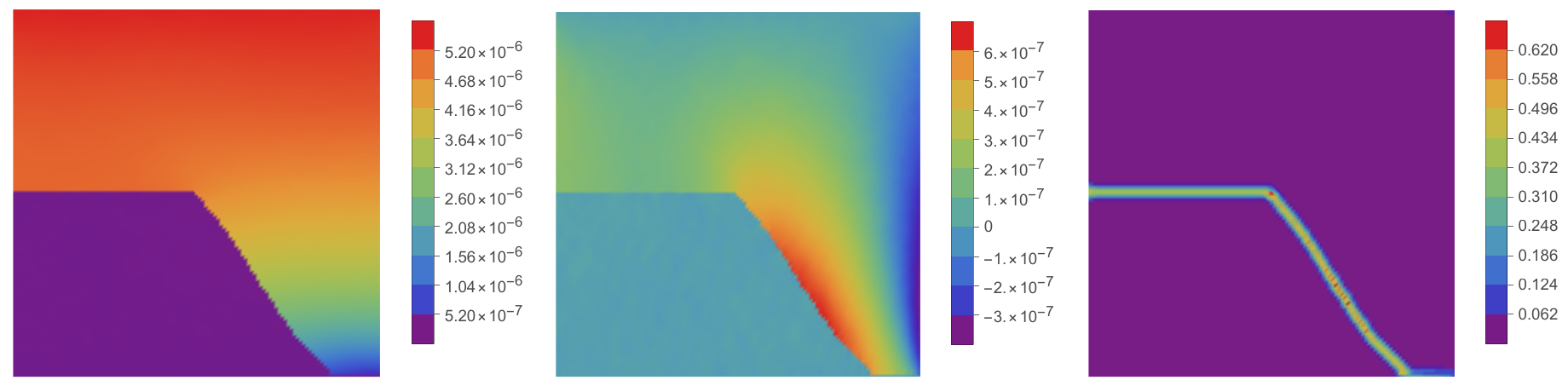}
	\caption{Single-edge-notched plate subjected to shear boundary condition: Displacement field in x-direction and y-direction and the damage distribution.}
	\label{fig:120UxUyDd28}
\end{figure}

\begin{figure}[htp]
\centering
\includegraphics[width=10cm]{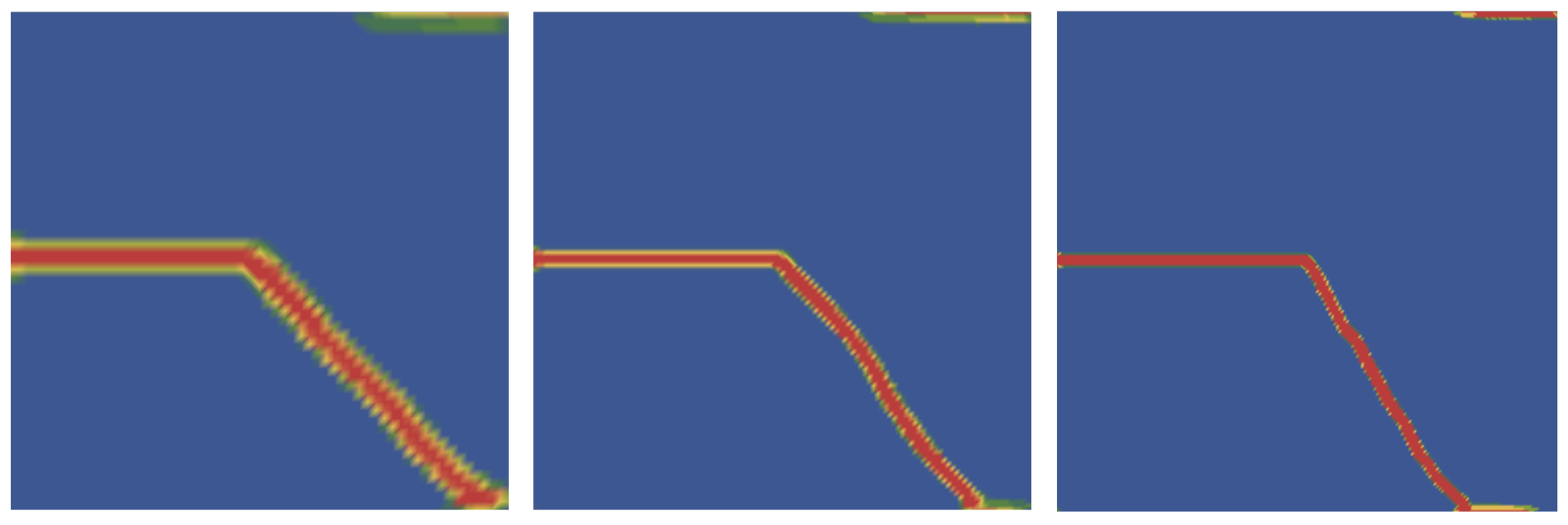}
\caption{Damage subjected to shear load.}
\label{fig:shearLoadDamage}
\end{figure}
The final result for shear tests with a discretization of $120\times 120$ is depicted in Figure \ref{fig:120UxUyDd28}.
In Figure \ref{fig:shearLoadDamage}, the damage patterns for different discretization subjected to shear loading condition are plotted. With finer discretization, the resolution of fracture becomes sharper.

\begin{figure}
	\centering
		\includegraphics[width=8cm]{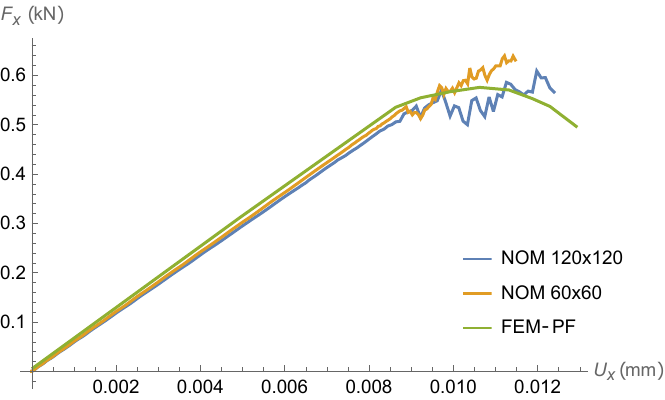}
	\caption{Single-edge-notched shear test: load curves for NOM and phase field by FEM.}
	\label{fig:shearLoadCurve}
\end{figure}
The graph depicting the displacement curve for the shear test has been illustrated in Figure \ref{fig:shearLoadCurve}. Prior to the initiation of the crack, the outcome obtained through the use of NOM is consistent with the finite element method outcome. The initiation of the crack occurred at a displacement of $u_x=0.009$ mm for both Finite Element Method and NOM. As the applied load is augmented, the process of bond cutting exhibits irregularities and the resultant reaction force undergoes oscillations. This phenomenon can be attributed to the misalignment between the particle distribution in the support and the surface of the crack. Additionally, it exposes the intricate stress condition resulting from the disturbance of fractured bonds. The fracture pattern observed is consistent with the results obtained through the finite element phase field approach. The current method exhibits a discrete feature, resulting in a less smooth crack surface compared to continuum methods like the phase field.

\subsection{Critical shear damage criterion}
\begin{figure}
	\centering
		\includegraphics[width=7cm]{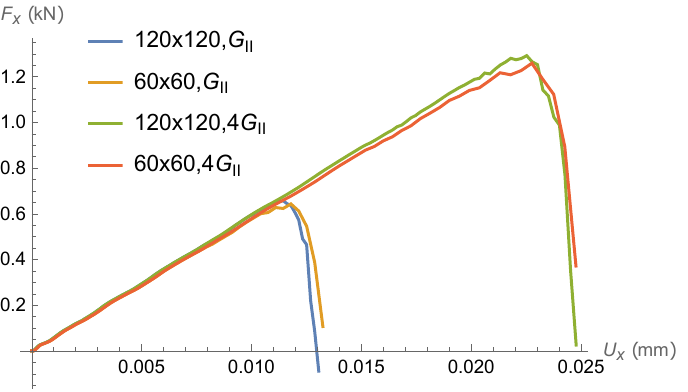}
	\caption{Single-edge-notched shear test: load curves for the case of $u_x:u_y=2:0$.}
	\label{fig:FxForceCurveShear}
\end{figure}
\begin{figure}
	\centering
		\includegraphics[width=7cm]{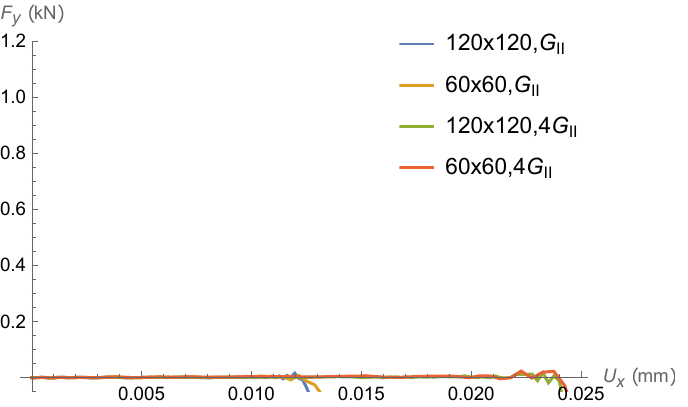}
	\caption{Single-edge-notched shear test: load curves for the case of $u_x:u_y=2:0$.}
	\label{fig:FyForceCurveShear}
\end{figure}
The material parameters remain consistent with those outlined in prior sections, with the exception of the energy release rate for mode II, which has been designated as $G_{II}=3\times 10^{-3}$ kN/mm. The determination of the critical shear stretch is conducted through the utilization of Equation \ref{eq:secrit}. The figures denoted as Figure \ref{fig:FxForceCurveShear} and Figure \ref{fig:FyForceCurveShear} respectively represent the resultant reaction forces in the $x$ and $y$ directions of the material points located at the uppermost section of the plate. Under pure shear boundary conditions, the force exerted in the $y$ direction is negligible in comparison to the force exerted in the $x$ direction. The magnitude of the peak reaction force exhibits a direct proportionality to the square root of the second mode of the energy release rate, denoted as $G_{II}$. In the scenario where the structure is discretized by a $120 \times 120$ grid and the critical energy release rate is selected as $4 G_{II}$, the maximum reaction force is observed to be $F_{x}^{max}=1.29$ kN at a displacement of $u_{x}=2.253\times 10^{-2}$ mm. This displacement corresponds to an external work of approximately $W_{ext}=\frac 12 F_{x}^{max} u_{x}=1.455 \times 10^{-2}$ J. The energy required for the formation of a fracture surface can be expressed as $2 \times (4 G_{II}) \times l_{crack}$, where $l_{crack}$ denotes the length of the crack and the factor of 2 accounts for the presence of two crack surfaces. Substituting $l_{crack}=0.5$ mm and $G_{II}=3\times 10^{-3}$ yields a value of $1.2 \times 10^{-2}$ J. The fracture energy exhibits a marginal reduction in comparison to the aggregate external work denoted by $W_{ext}$. The outcome is deemed rational as the overall energy is composed of specific proportions of both kinetic and strain energy.

\begin{figure}
	\centering
		\includegraphics[width=5cm]{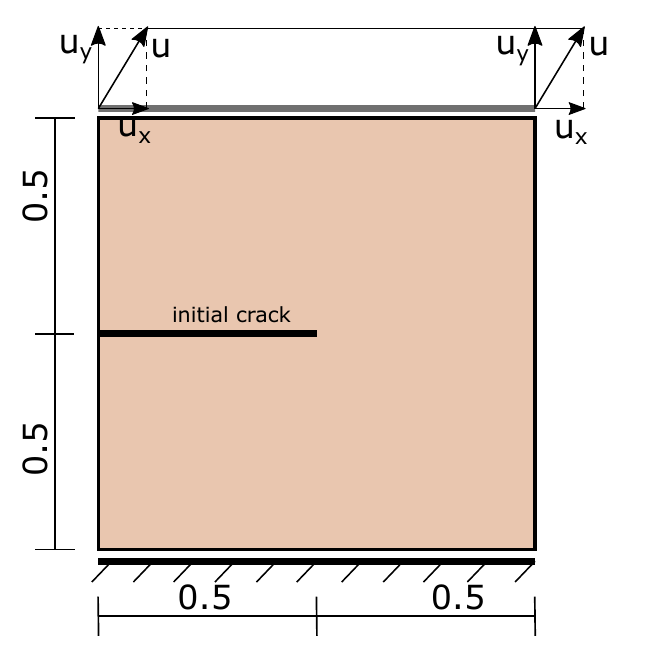}
	\caption{Single-edge-notched test based on shear damage criterion: displacement boundary. }
	\label{fig:notchPlateOblique}
\end{figure}
\begin{figure}
	\centering
		\includegraphics[width=16cm]{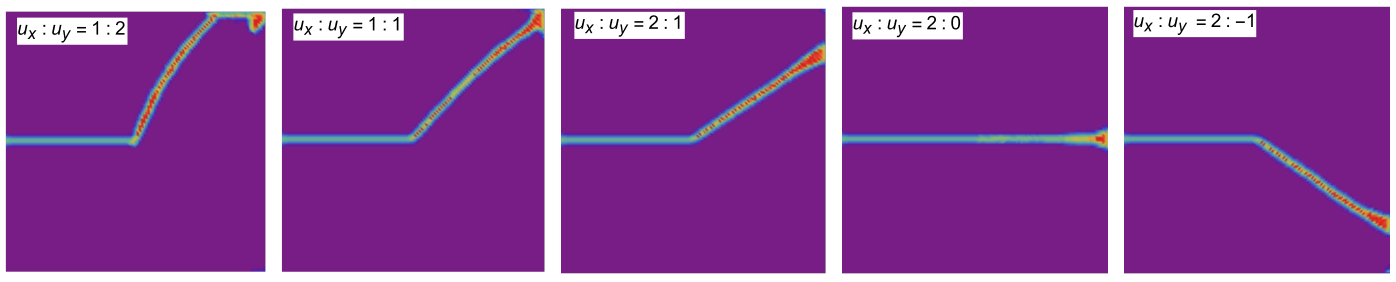}
	\caption{Single-edge-notched shear strain: fracture patterns on different displacement boundaries.}
	\label{fig:ShearFractureModes}
\end{figure}

Additionally, we assess the impact of the loading angle as shown in Figure \ref{fig:notchPlateOblique}. Through manipulation of the $u_x:u_y$ ratio, various modes of shear crack can be observed, as depicted in Figure \ref{fig:ShearFractureModes}. Remarkably, the direction of the crack path exhibits significant proximity to the direction of displacement. The crack paths in the scenario where the ratio of $u_x$ to $u_y$ is 2:1 and the scenario where the ratio is 2:-1 exhibit horizontal line symmetry. The application of the critical shear strain damage rule results in the automatic identification of the direction of maximal shear strain and the consistent formation of a shear crack path, leading to a sufficient inference.

%\end{frame}
\subsection{Kalthoff-Winkler experiments}\label{sec:wk}
\begin{figure}[htp]
	\centering
		\includegraphics[width=4cm]{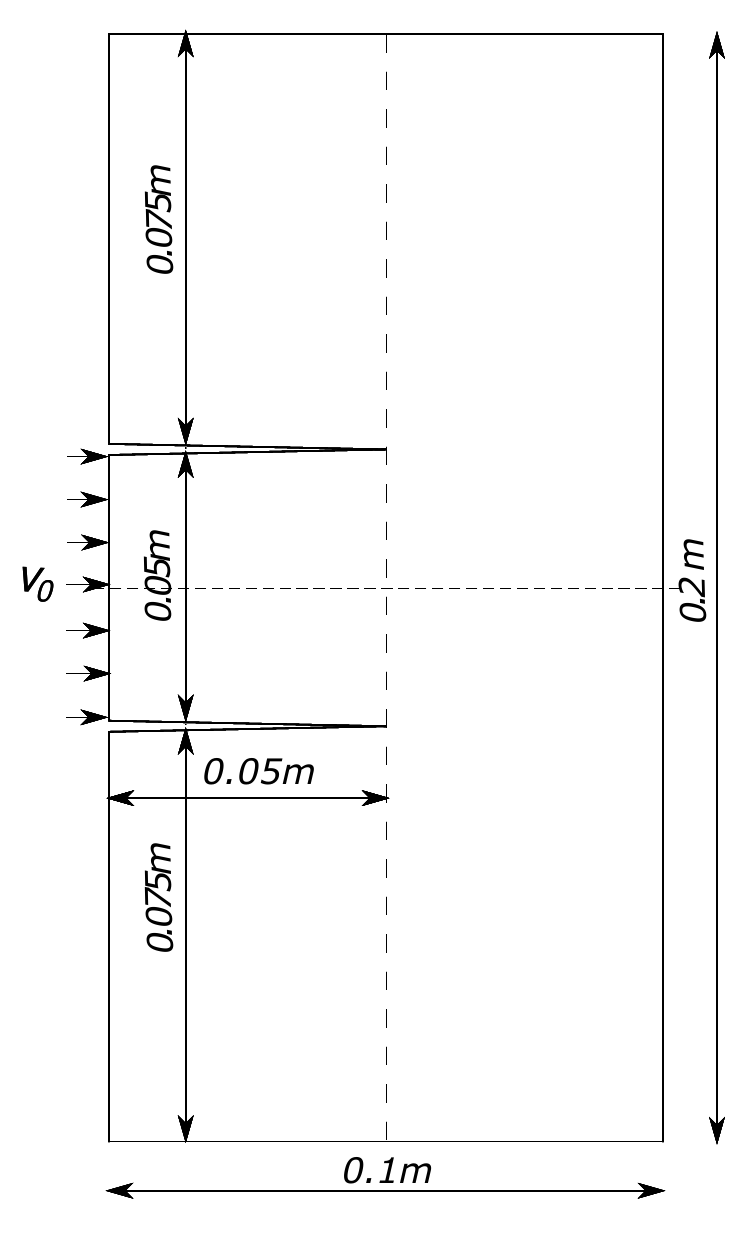}
	\caption{Setup for the Kalthoff-Winkler experiment.}
	\label{fig:kwsSetup}
\end{figure}
The Kalthoff-Winkler experiment has been widely recognized as a classical benchmark problem in the field of dynamic fracture modeling \cite{kalthoff1988failure, Li2002, Belytschko2003, Song2006}. The fracture may exhibit either brittle or ductile behavior, depending on the varying impact velocities. At low impact velocities, the dynamic brittle fracture spreads from the crack tip at an approximate angle of 70$^{\circ}$ relative to the orientation of the initially horizontal crack. Upon further increase in impact velocity, a ductile failure, also known as shear fracture, is observed, accompanied by the formation of a shear band. The plate's dimensions are 0.2 meters by 0.1 meters, as depicted in Figure \ref{fig:kwsSetup}. The given values for the material parameters are as follows: Young's modulus, Poisson's ratio, and critical energy release rate  $E=190\mbox{ GPa} ,\nu=0.3, G_c=2.4\times 10^4\mbox{ J/m} ^2$. With consideration for symmetry, only half of the plate is modeled. The plate has been discretized into a grid of 200 by 200 particles. The chosen value for the support radius is $l=3\Delta x$ meters. The initial crack is indicated by altering the adjacent neighbors in the support. The maximum number of neighboring particles of each particle is selected as 28. The critical normal strain criterion is utilized to investigate brittle failure under low impact velocity. The velocity imposed on the impact surface of the plate undergoes a step change from an initial value of zero to a final value of $v_y=20$ m/s within a duration of $10^{-7}$ s and subsequently remains constant   \cite{miehe2015phase}. In the case where shear fracture occurs at an elevated impact velocity of $v_y=39$ m/s, the critical shear damage criterion is utilized. Additionally, the energy release rate of mode-II fracture is designated as $G_{II}=4 G_c$.

\begin{figure}
	\centering
		\includegraphics[width=16cm]{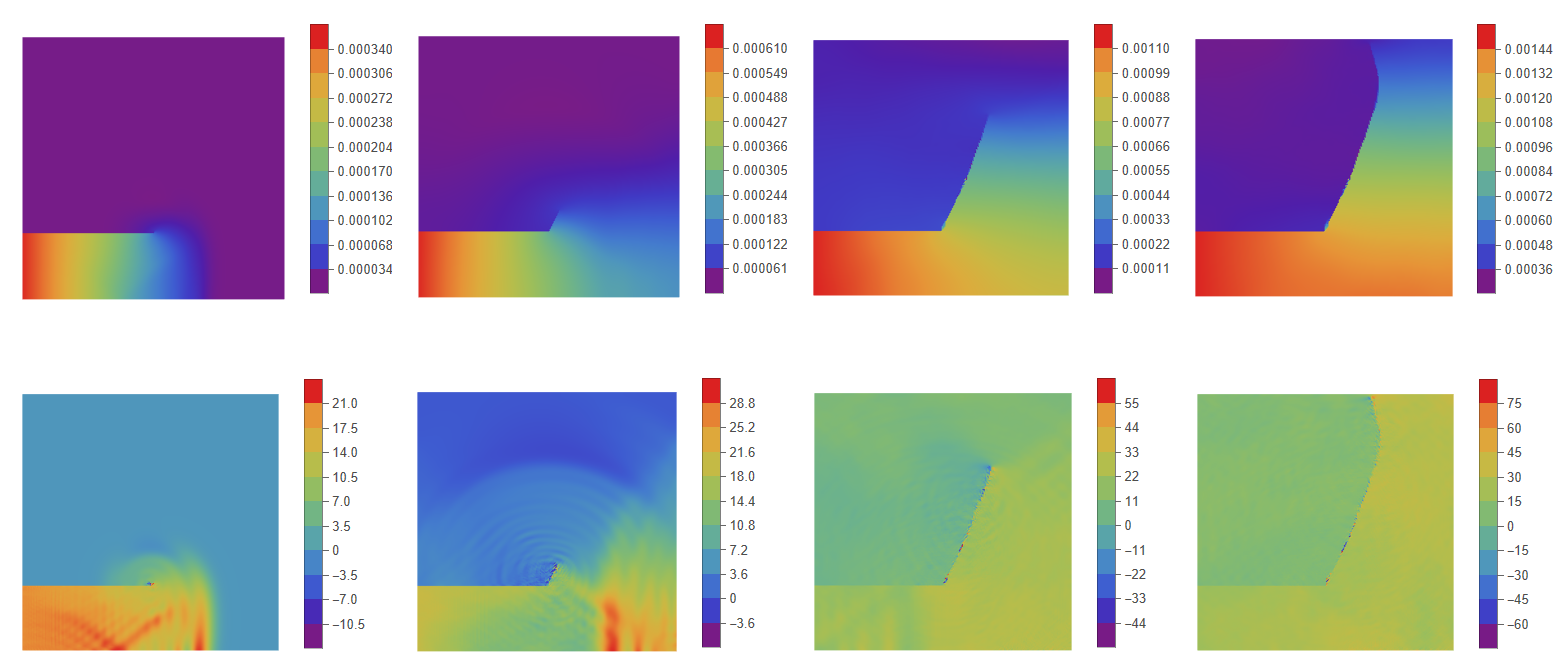}
	\caption{Kalthoff-Winkler test $v_y=20$ m/s: $u_x$ (first row) and $v_x$ (second row) at times $(24  \mu s, 33.7  \mu s, 62.6  \mu s, 82 \mu s)$.}
	\label{fig:KWtensileUxVx}
\end{figure}

\begin{figure}
	\centering
		\includegraphics[width=5cm]{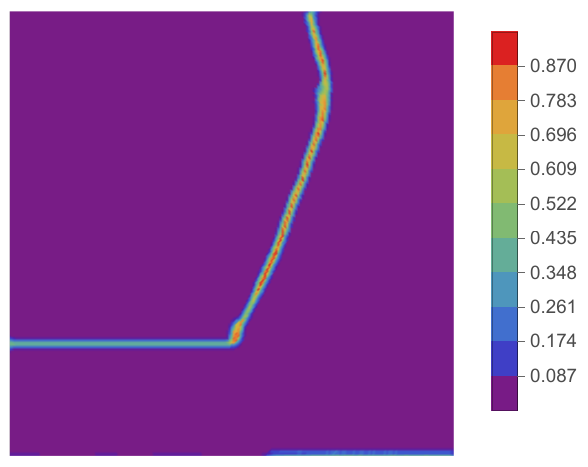}
	\caption{Kalthoff-Winkler test $v_y=20$ m/s: tensile fractures.}
	\label{fig:KWtensileCrack}
\end{figure}

\begin{figure}
	\centering
		\includegraphics[width=16cm]{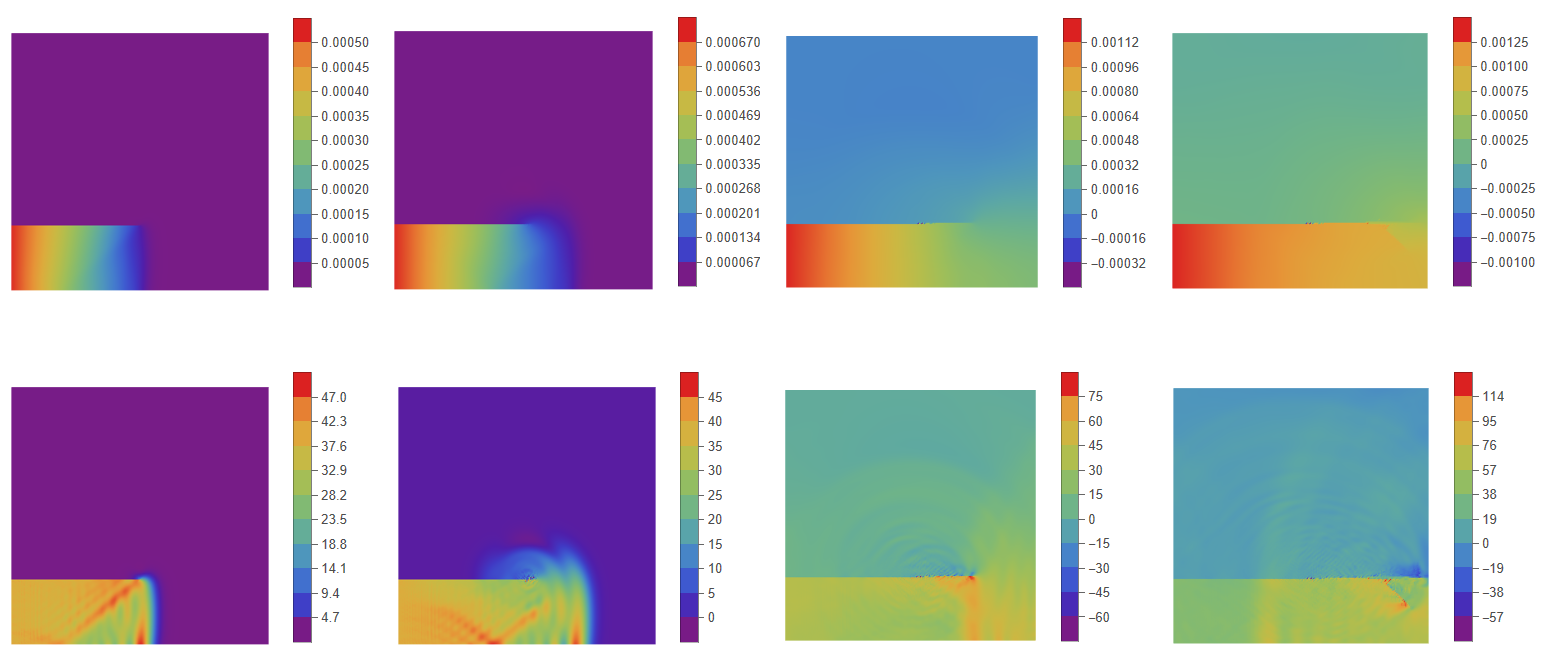}
	\caption{Kalthoff-Winkler test $v_y=39$ m/s: $u_x$ (first row) and $v_x$ (second row) at times $(14.4  \mu s, 19.3  \mu s,  33.7  \mu s,43.4 \mu s)$.}
	\label{fig:KWshearUxVx}
\end{figure}

\begin{figure}
	\centering
		\includegraphics[width=5cm]{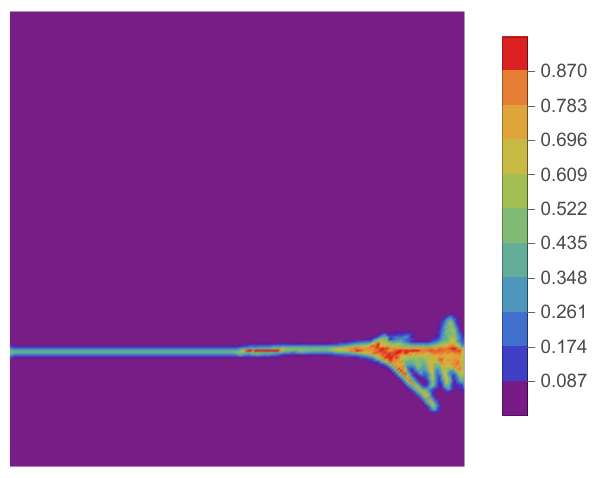}
	\caption{Kalthoff-Winkler test $v_y=39$ m/s: shear fractures.}
	\label{fig:KWshearCrack}
\end{figure}

Figure \ref{fig:KWtensileUxVx} illustrates the displacement and velocity fields in the $x$-direction at various time intervals under low impact velocity conditions. The data indicates that the crack commenced at a time of $24 \mu s$ and terminated at $82\mu s$. The breaking of bonds in the vicinity of the crack tip results in noticeable oscillations in velocity. The ultimate trajectory of the tensile crack is depicted in Figure \ref{fig:KWtensileCrack}. Figure \ref{fig:KWshearUxVx} displays the displacement field and velocity for the impact velocity of greater magnitude. The initiation of shear crack propagation occurs at a time of $14.4 \mu s$, and it proceeds in the same direction as the original crack. During the concluding phases, the division of shear fractures into crack branches can be observed, as depicted in Figure \ref{fig:KWshearCrack}.
%{14.4, 19.3, 33.7, 43.4}


\section{Conclusions}

The present study introduces several bond-based models for solids, thin plates, and gradient solids in diverse dimensional spaces. The primary objective is to establish a bond force model that is solely dependent upon bond deformation, with the local theory being restored through collective deformations via the application of an energy equivalence principle and the assumption of a fully symmetric support region. The bond-based NOM incorporates a weight function to establish the bond forces. The introduction of a bent bond enables the definition of nonlocal curvature and nonlocal moment, which account for the bending effect resulting from curvature. The symmetrical definition of the bent bond encompasses three points. The thin plate model utilizing bonds has been demonstrated to possess a constraint on its Poisson's ratio. A bond-based gradient elasticity model is derived from the principle of equivalence of the gradient deformation energy between local and nonlocal settings.

The bond-based elasticity takes into consideration the normal deformation and shear deformation in a bond without being constrained by Poisson's ratio. The regulation of the distribution of nonlocal bond strain energy is achieved through the use of a weight function. Simultaneously, a damage model is suggested for deformed bonds, whereby the bond strength is reduced upon attainment of the bond strain or bond curvature threshold value. This configuration offers a straightforward guideline for the localization of strain without the need to sever the bond. Furthermore, a plasticity model is formulated by utilizing the incremental deformation of a bond.

The work provides several numerical instances, such as a simply supported beam and a two-dimensional solid plate exhibiting shear or tensile damage patterns. While the numerical examples in this study employ explicit time integration, the implicit implementation is straightforward for static problems. It is possible to compute the second variation for each bond element and transform the tangent stiffness matrix from the local coordinate system to the global coordinate system. Finally, a straightforward guideline is suggested for determining the critical normal and shear strains for bond cutting in models of tensile and shear fracture, respectively. This guideline is both computationally stable and straightforward to execute, and it yields outcomes that are comparable to those obtained through the use of the phase field approach.

\section*{Acknowledgments}
The first author gratefully acknowledges the financial support from the EU project entitled ''Computational Modeling, Topological Optimization, and Design of Flexoelectric Nano Energy Harvesters'' (ERC COTOFLEXI 802205).

\bibliographystyle{unsrt}
\bibliography{nombond}
\end{document}